\documentclass[12pt,preprint]{aastex}
\usepackage{graphicx}
\usepackage{natbib}
\usepackage{changepage} 
\usepackage{comment}
\def\sax{{\em BeppoSAX\/}}

\def\ergcm{\mbox{ erg cm$^{-2}$}}
\def\ergcms{\mbox{ erg cm$^{-2}$ s$^{-1}$}}
\makeatletter
\def\@cite#1#2{(#1\if@tempswa , #2\fi)}
\def\preprint{preprint}   \newif\ifPreprintMode
\ifx\preprint\revtex@genre\PreprintModetrue\else\PreprintModefalse\fi
\makeatother

\begin{document}

\title{The Gamma--Ray Burst catalog obtained with the Gamma Ray Burst Monitor
aboard \sax}

\author{F.~Frontera\altaffilmark{1,2},
C.~Guidorzi\altaffilmark{3},
E.~Montanari\altaffilmark{1,4},
F.~Rossi\altaffilmark{1},
E.~Costa\altaffilmark{5},
M.~Feroci\altaffilmark{5},
F.~Calura\altaffilmark{6},
M.~Rapisarda\altaffilmark{7},
L.~Amati\altaffilmark{2},
D.~Carturan\altaffilmark{1},
M.~R.~Cinti\altaffilmark{5},
D.~Dal Fiume\altaffilmark{2},
L.~Nicastro\altaffilmark{2},
M.~Orlandini\altaffilmark{2}
}

\altaffiltext{1}{Universit\`a di Ferrara, Dipartimento di Fisica, Via Saragat 1, 
44100 Ferrara, Italy; email: frontera@fe.infn.it}
\altaffiltext{2}{INAF/Istituto di Astrofisica Spaziale e Fisica Cosmica,
Bologna, Via Gobetti 101, 40129 Bologna, Italy}
\altaffiltext{3}{INAF, Osservatorio Astronomico di Brera, Via Bianchi 46, 23807 Merate,
Italy}
\altaffiltext{4}{Istituto IS Calvi, Finale Emilia (MO), Italy} 
\altaffiltext{5}{INAF/Istituto di Astrofisica Spaziale e Fisica Cosmica,
Roma, Via del Fosso del Cavaliere 100, 00133 Roma, Italy}
\altaffiltext{6}{Dipartimento di Astronomia,  Universit\`a di Trieste,
Via G. B. Tiepolo 11,  34131 Trieste, Italy}
\altaffiltext{7}{ENEA Divisione Fusione, Centro Ricerche di Frascati,
  CP 65, 00044 Frascati, Rome, Italy}

%

\begin{abstract}
We report on the catalog of Gamma--Ray Bursts (GRBs) detected with the Gamma Ray Burst
Monitor aboard the \sax\ satellite. It includes 1082 GRBs with 40--700 keV fluences in
the range from $1.3\times 10^{-7}$ to $4.5\times 
10^{-4}$~erg~cm$^{-2}$, and 40--700 keV peak fluxes from $3.7\times 10^{-8}$
to $7.0\times 10^{-5}$~erg~cm$^{-2}$~s$^{-1}$. We report in the catalog some relevant
parameters of each GRB and discuss the derived statistical properties.
\end{abstract}


\section{Introduction}
\label{intro}
In the last ten years a big step forward has been accomplished in the
Gamma Ray Burst (GRB) astronomy. A key role in this advancement has been played by
the \sax\ satellite \citep{Boella97a}. Its capability of promptly and
accurately localizing GRBs and the possibility of following them up with highly sensitive
X--ray telescopes on-board allowed the discovery of the 
X--ray afterglow emission and, with ground telescopes, of the 
optical/radio afterglow from GRB sources \citep{Costa97,Vanparadijs97,Metzger97,Frail97}. 
In this way the distance scale of long 
($>$1~s) GRBs could be eventually established and important information 
on the GRB sources, their emission mechanisms and their environments could
be derived.

The two \sax\ instruments that allowed the prompt and accurate GRB localization were 
the Gamma Ray Burst Monitor (GRBM) and the two Wide Field Cameras (WFC), the first
with the role of recognizing the occurrence of a GRB and second with the role
of localizing it in the case the event direction occurred in their field of view
($40^\circ\times 40^\circ$ full width at zero response) and was above the
instrument sensitivity in their operative
energy range (2--28 keV)\citep{Jager97}. 
Only a handful of GRBs (50, see Frontera 2004,\nocite{Frontera04c}) was detected with both
instruments, while the GRBM alone detected 1082 GRBs, a number similar to that
of GRBs included in the third BATSE (3B) burst catalog~\citep{Meegan96}. In this paper
we present the catalog of these GRBs with their main observational and statistical
properties.

\section{Instrumentation}
\label{instru}
\begin{figure}
\begin{center}
\includegraphics[angle=0,width=0.9\textwidth]{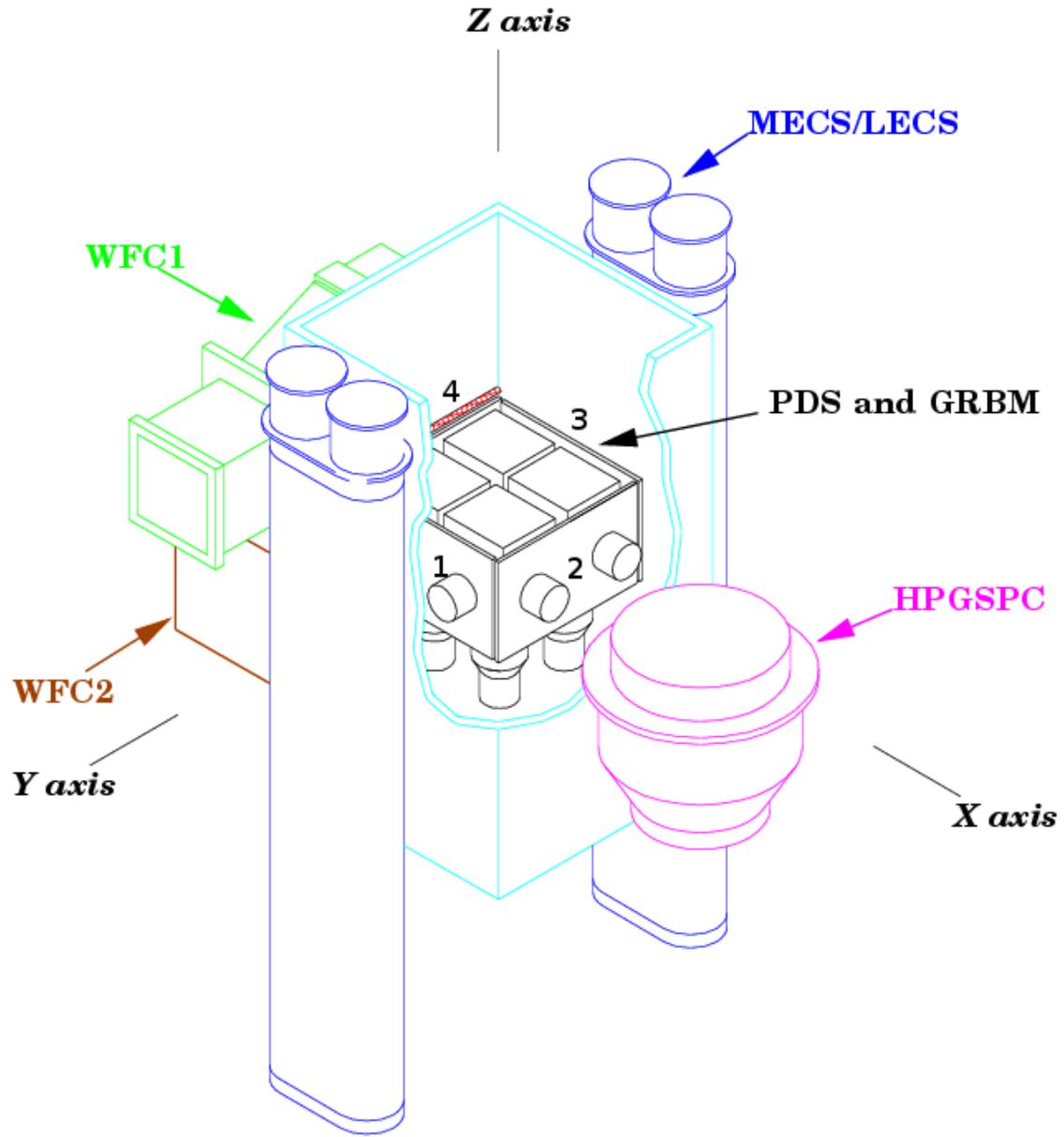}
\end{center}
\caption{The \sax\ payload, in which the location of the GRBM units is shown.}
\label{f:sax-payload}
\end{figure}

The \sax\ GRBM was part of the high energy experiment PDS (Phoswich Detection System,
\citet{Frontera97}), being its anti-coincidence (AC) shield.
It was composed of four independent detection units, made of slabs of 
CsI(Na)scintillators, forming the four sides of a square box 
(see Fig.~\ref{f:sax-payload}). Each
slab was 1 cm thick and had a geometric area of 1136 cm$^2$, with an open
field of view. The light scintillation produced in each unit was viewed by
two photo-multipliers. A calibration source, made of Am$^{241}$, embedded in a NaI(Tl)
crystal, allowed to monitor the instrument gain. The gain could
be varied by changing the HV supply of each GRBM unit. 

The electronics associated to the GRBM was a simple spectroscopic chain in which
the signals from each of the four detection units  between two analog thresholds, after
being multiplexed, were  analog-to-digital converted by an ADC. 
By telecommand the low level threshold (LLT) could be varied in the nominal energy
range from 20 to 90 keV (16 steps), while the upper level threshold (ULT)could be
varied between 200 and 700 keV (8 steps).
The signals detected by each unit between LLT and ULT (good events) and 
above the AC threshold (ACT) were continuously counted with a 1 s integration time
and stored in the on-board memory. The ACT could be selected between 100 and 300 keV 
(8 steps).
The GRBM electronic unit included an on-board GRB trigger (see \S~\ref{trigger}). 
If the on--board trigger condition was satisfied for at least two detection units, 
then the following high time resolution profiles for each of the GRBM units 
were stored on board and then transmitted to ground:
\begin{itemize}
\item  7.8125 ms time profiles from the trigger time back to 8 s before;
\item 0.48828125 ms time profiles from the trigger time to 10 s later;
\item 7.8125 ms time profiles from 10 s after the trigger time for 88 s.
\end{itemize}
In addition to these data, for each detection unit, the following data were 
continuously transmitted:
\begin{itemize}
\item  1 s count rates between LLT and ULT;
\item 1 s count rates above ACT (in phase with the previous ones);
\item 240 channel spectra of the signals between LLT and ULT,
integrated over 128 s (synchronous integration, in phase with the 1~s count rates).
\end{itemize}
Further details about the GRBM instrument can be found elsewhere \citep{Amati97,Feroci97,Costa98}.

\subsection{On--board and on--ground GRB trigger logic}
\label{trigger}

The GRBM electronic unit included an on-board GRB trigger logic (OBTL) that worked 
in the following way. For each GRBM unit and for the events with amplitudes between 
LLT and ULT with a binning time
of 7.8125 ms, a moving average on a  Long Integration Time (LIT) was
continuously computed. The LIT was adjustable from the ground between 8 and 128 s.
The counts integrated over a Short Integration Time (SIT), also adjustable between
7.8125 ms and 4 s, were thus compared with the LIT moving average normalized to the
same integration  time. If the count difference exceeded $n$ ($n$ = 4, 8, or 16)
times the Poissonian
standard deviation $\sigma$ of the mean count rate, then the trigger condition
for that unit was satisfied. As seen above, the GRB trigger was activated when
the trigger condition was satisfied for at least 2 detection units.

As discussed below, the OBTL was not suitable to distinguish between
true GRBs and many other spurious  events, like the crossing of high energy 
charged particles, atmospheric phenomena, etc. In addition, it was not able
to detect weak events, that were recognized as GRBs only by visual inspection of the
1 s ratemeters; also it was not operative during the transmission of the
high time resolution data to the on-board tape recorder. In the latter case, only
a flag about the occurrence of a burst--like event was transmitted.

In order to reject spurious events and/or to increase the GRB detection 
rate, we implemented a ground software \citep{Guidorzi01} which was capable to perform a prompt
analysis of the GRBM data transmitted each orbit when the satellite was visible 
by the Malindi (Kenya) ground station. The SW trigger conditions (SWTC) 
were the following. As in the case of the OBTL, for each GRBM detection unit 
the count level in the 40--700 keV and $>100$~keV nominal energy bands
was continuously monitored by performing a moving average integrated over 100 s time interval. 
A SW trigger was generated each time the counting rate either in both energy bands 
for at least two detection units, or in both energy bands for only one unit whose
axis was directed toward the sky, or only in the 40--700 keV
band for at least three detection units, was $n \sigma$ higher than the last 
estimate of the moving average, with $n = 6$ in the 40--700 keV band, and $n = 3$
in the $>100$~keV channel.
The condition for the $>100$ keV channel was of key importance for recognizing a
true GRB event and to reject spikes due to high energy particles, whose energy release 
in the GRBM detectors showed a soft spectrum.

\section{GRBM response function}

The determination of the GRBM response function, both in direction and energy,
was performed with Monte Carlo techniques and tested with  ground 
calibrations performed before the \sax\ launch and with various observations of the
the Crab Nebula. It was also cross-calibrated with the BATSE experiment
\citep{Fishman94}.

The GRBM response function derived via Monte Carlo code \citep{Rapisarda97} 
made use of the Monte Carlo N-Particle Transport Code (MCNP), version 4.2, 
released by the Los Alamos National Laboratory (LANL) (for the current 
MCNP version see \citet{Brown02}).
The code allows to transport photons, neutrons and electrons from 1 keV to
100 MeV through matter. Every kind of three--dimensional geometry could be described
by defining separate cells with composition and density chosen by the user.
Photon interactions were treated very accurately and the entire satellite was
modeled with a high degree of detail.

The model was tested using the GRBM ground calibrations performed before the
flight when the instrument was integrated in the spacecraft \citep{Amati97}. Preliminary
results of these tests were reported \citep{Calura00}.

The response function, obtained with the Monte Carlo code, was determined as a 
function of the energy and  
direction of the incident photons. Using as reference frame the satellite local
frame with equatorial plane perpendicular to the axis of the Narrow 
Field Instruments
($z$ axis), and using as polar axis the $z$ axis, a grid of 576 directions was 
determined: 36 azimuthal angles $\phi_k$ in the range 0 to 360 degrees, in which
the zero value corresponds to the azimuth of the axis of the detection unit \#2,
and 16 polar angles $\theta_j$ in the range $-70<\theta<+80$ degrees. Ten different 
intervals of the photon energy $E_n$ ($n = 1, 11$) in the 30 to 1000 keV were also simulated
with logarithmic steps. For each grid point we determined the expected number 
of counts $N_i(E_n, \theta_j, \phi_k)$  detected 
in the  40--700 keV and $>$ 100 keV channels by each of the 4 detection units
($i = 1, 2, 3, 4$ for detections by the units \#1, \#2, \#3, \#4 in the 40--700 keV 
channel, and $i = 5, 6, 7, 8$ for detections by the same units in the $>100$ keV channel),  
for an input photon of energy $E_n$ incident from the direction $(\theta_j, \phi_k)$. 

A further refinement of the
grid was obtained by interpolating with cubic splines $N_i(E_n, \theta_j, \phi_k)$ in both 
energy and direction. The grid energy was refined up to  1 keV steps and the direction
$(\theta_j, \phi_k)$ was refined up to $(1^\circ, 1^\circ)$ steps. 
Assuming a photon spectrum $I(E)= K f(E,{\bf \alpha})$, where
{\bf $\alpha$} are the model parameters, for a photon beam coming from the direction 
$(\theta_j,\phi_k)$ we 
expect to detect from each  detection unit  the following  count rate in the above 
energy channels by each detection unit: 
\begin{equation}
C_i(\theta_j,\phi_k,{\bf \alpha}, K) = \sum_{n} \Delta E_n K f(E_n,{\bf \alpha}) 
N_i(E_n, \theta_j, \phi_k)
\label{exp_counts}
\end{equation}

The best estimate of the GRB arrival direction $(\theta,\phi)$ and of the parameters 
{\bf $\alpha$} of the assumed photon spectrum is determined by minimizing the following $\chi^2$ 
statistics (for a power-law photon spectrum, see \citet{Calura00}):
\begin{equation}
\chi^2 (\theta_j,\phi_k, K, {\bf \alpha}) = \sum_{i=1}^{8}\ \frac{1}{{\sigma^2_i}}\  
\left( n_i - \frac{n\,C_i(\theta_j,\phi_k, K, {\bf \alpha})}
{C(\theta_j,\phi_k)}
\right )^2
\qquad
\label{chisq_min}
\end{equation}

where $n_i$ is the measured counting rate measured by the unit $i$ as above defined, 
$\sigma_i^2$ is the variance of $n_i$, $n = 
\sum_{i=1}^{8} n_i$, and $C(\theta_j,\phi_k) = \sum_{i=1}^{8}C_i(\theta_j,\phi_k, K, \alpha)$.

Actually, given that we had only two energy channels, we were constrained to use as input model
$f(E,{\bf \alpha})$ a simple power--law ($E^{-\Gamma}$). The best estimate of the parameters 
$(\theta, \phi, K,\Gamma)$ was that which minimized the $\chi^2$ in the eq.~\ref{chisq_min}. 
In the case that more solutions were found for $(\theta, \phi)$, no localization or spectrum is
included in the catalog.   

\section{In-flight performance}
  
\subsection{In-flight settings}
\label{setting}
The values of the GRBM thresholds and trigger parameters were set
and varied during the \sax\ mission (inclination orbit of 3.9$^\circ$,
initial altitude of 600 km) according to Table~\ref{t:settings}. 
The initial low energy threshold (32.5 keV) was risen to 42.5 keV in October 1996
in order to suppress the spiking noise and to decrease the background level, 
without a significant
change of the source signal given the small transparency of the GRBM at
low energies due to payload materials in its Field of View (FOV) 
(see Fig.~\ref{f:sax-payload}). 
With the new low-energy threshold and  the new setting of the GRB trigger 
parameters (see Table~\ref{t:settings}), about 12 triggers per day occurred
($\sim$0.8 triggers per orbit), most of which were false. The false triggers
were due to correlated events in two contiguous detection units. These events
were mainly due to charged particles which crossed two detection units. Indeed we
found that false triggers were composed of fast events (less than 10 ms duration)
with slightly different time profiles in two contiguous detection units.

The decrease of the LIT from 128 s to 32 s in November 1996 (see 
Table~\ref{t:settings}) was motivated by the need of a correct evaluation 
of the background level when the satellite was approaching to the South Atlantic
geomagnetic Anomaly (SAA). Indeed we found that 10 minutes before SAA the 
background significantly increased below 100 keV, sometimes dramatically. 
One of these rapid variations occurred during the orbit in which we detected
the famous GRB event of 28 February 1997~\citep{Costa97}. We called these 
rapid background variations pre--SAA effects \citep{Feroci97}.

The last change of the GRB trigger parameters, in particular of the SIT,
was performed in 2002, a few months before the end of the \sax\ operative
life, in order to make the GRB trigger more sensitive to short GRBs.

\subsection{Background estimate and its subtraction}

The GRBM was nominally performing since the satellite launch \citep{Feroci97}. 
The background level 
along the orbit, outside the region near the SAA, remained very stable, 
with an average variation  $\le 10$\%. Because of 
the background stability, 
the background subtraction for the GRB data analysis was in general not critical, except 
for very weak events occurring close to the SAA and, for 
spectral evolution analysis, during the initial onset and tails of weak events. 
For the ratemeters data, the background during the event was estimated by 
interpolating the data taken $\sim 250$~s before and after the event with a 
polynomial, the order of  which was chosen on the basis of the local
background variation level. 
In the case of the 240-channel spectra, polynomials were used to interpolate the 
background trend in different energy ranges using 3 (or more) packets (each one covering 
128 s) before and after the packet (o more packets) containing the event spectrum.

\section{In flight test of the GRBM response function}

After background subtraction, GRBM data were analyzed using the 
response matrix described above, which had also been converted to 
FITS format to be used with standard spectral analysis software packages like 
XSPEC. 

The goodness of the response matrix was verified in flight with the Crab Nebula
observed via source occultation by the Earth \citep{Guidorzi98} and with cross--checks 
with BATSE results obtained for a GRB selected sample by \citet{Kaneko07}.

The Crab flux and spectrum were derived using both the 2-channel ratemeters 
and the 240-channel spectra. We found the spectral parameters consistent 
with the corresponding values found with other experiments in hard X--/soft gamma--rays, 
i.e. a photon index of about 2.2 and  a 100 keV flux density of about 
60 $\times 10^{-5}$ photons~cm$^{-2}$~s$^{-1}$~keV$^{-1}$, with reduced $\chi^2$ 
values of about 1.3 for about 13 d.o.f. By 
fixing the photon index at the commonly adopted X-rays value of 2.1, we obtain a 
normalization at 1 keV of $9.64\pm 0.49$ photons~cm$^{-2}$~s$^{-1}$~keV$^{-1}$ , 
which is fully consistent with the classical value of 9.7 \citep{Toor74}. 

Cross-checks with BATSE results were performed with a sample of 46 strong GRBs detected 
by both experiments and arriving from various directions. The results 
show that the ratio of the 40--700 keV fluences derived from the GRBM response function 
with those derived using spectral law, parameters and time integration given by \citet{Kaneko07},
is distributed according a Gaussian centered at about 0.8 with a standard deviation of 0.3. 
The deviation of the ratio mean value from 1 and the distribution spread  are a likely consequence 
of the uncertainties in the response functions and GRB directions, and of the differences 
in integration time and spectral models adopted by us and by the above authors.  

The uncertainty in the knowledge of the GRBM response 
function is on average of the order of 10\%. This uncertainty is mainly due to the error
in the knowledge of the flux of the calibration sources and to the errors 
in the calibrations data. 
This uncertainty was added in quadrature to that on the GRB count spectra before 
performing the fits.

\section{Sky exposure}

\begin{figure}
\begin{center}
\includegraphics[width=0.9\textwidth]{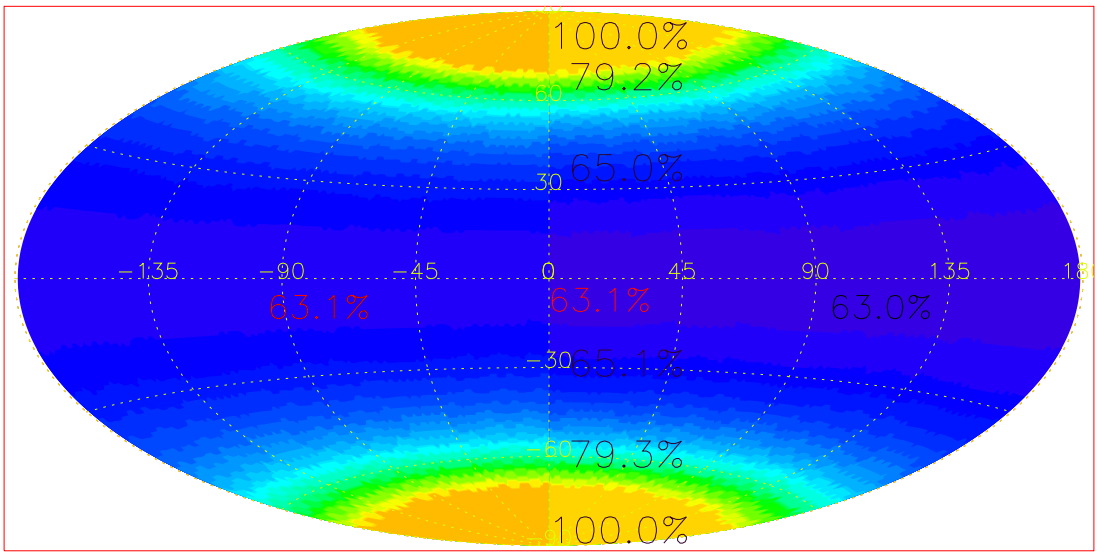}
\end{center}
\caption{Fraction of time in which a given sky direction, 
in celestial coordinates, was exposed to the GRBM.} 
\label{f:sky_exposure}
\end{figure}

\begin{figure}
\begin{center}
\includegraphics[angle=0,width=\textwidth]{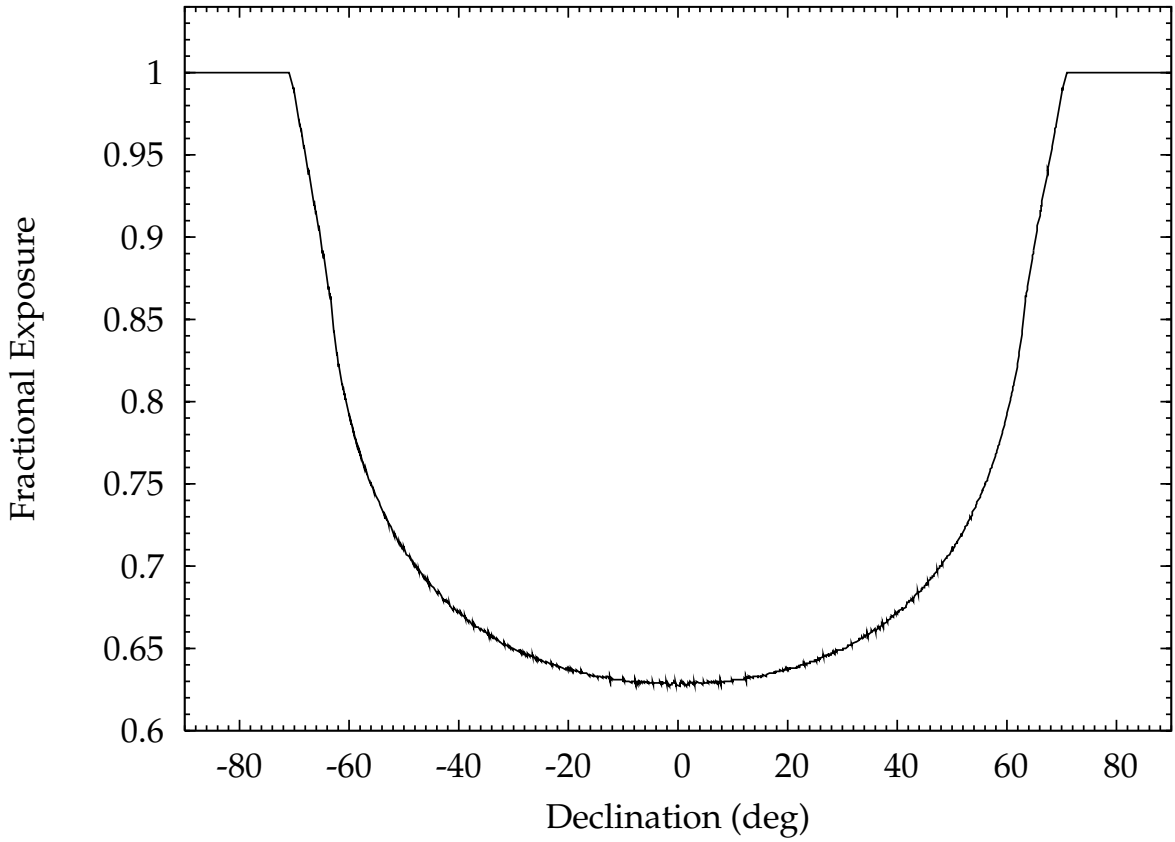}
\end{center}
\caption{Fraction of the GRBM exposure time versus celestial declination
after averaging the sky exposure over the right ascension.
} 
\label{f:dec_exposure}
\end{figure}

The GRBM sky exposure, that is the fraction of time above the horizon in which a given sky 
direction
was exposed to the GRBM, is shown in Figure~\ref{f:sky_exposure}, in celestial coordinates.
As can be seen, the celestial poles were never blocked by the Earth and thus they were
continuously visible by the GRBM, while the other sky directions, due to the
\sax\ orbit, were monitored for a shorter time. The dependence of the
sky exposure on the right ascension was very small, while that on the declination
is shown in Fig.~\ref{f:dec_exposure}, after averaging the sky exposure over the right 
ascension.

\section{The Catalog}
\label{s:cat}

In Table~\ref{t:cat} we show the GRBM catalog of GRBs. A preliminary version of the 
catalog can be found elsewhere \citep{Guidorzi02}, where the peak fluxes and fluences were 
given in counts (no correction for efficiency). The catalog includes 1082 events.
The parameters associated to each event include the best GRB direction 
in Galactic and equatorial coordinates, two different estimates of GRB time duration, i.e., 
the classical $T_{\rm 90}$ (time  during which the burst integrated 
counts increase from 5\% to 95\% of the total counts, Kouveliotou et al. 1993 \nocite{Kouveliotou93}) and the
time duration $T_{\rm det}$ (see below),
the integrated time $T_{\rm a}$ during which the burst count rate is detected
above a $2\sigma$ level, the 40--700 keV fluence, peak flux, the spectral hardness, the GRBM units used 
for the parameters determination, the binning factor with respect to the default count accumulation 
time (1 s for long GRBs, 7.8125~ms in the case of short GRBs). 
For different reasons not all the parameters were determined for the entire GRBs sample
(see details in Table~\ref{t:info}).
Details on the data reported and on methods adopted for the derivation of the parameters 
reported in the catalog are given below.

\subsection{GRB coordinates}

For each event, we report the most accurate equatorial (at the
epoch 2000.0) and Galactic coordinates available. 
When a GRB was detected by more than one experiment, 
the more plausibly precise localization is given. When the GRB was detected 
by one of the WFCs on
board \sax\, other possible detections are ignored. For the
GRBs detected by only GRBM and for which the localization procedure has
given a unique position, the GRBM direction estimate is reported. In
the  column {\em CAT} of the catalog we report the instrument or mission that provided
the best GRB coordinates.

\subsection{Detection units used for the parameter derivation}

For each GRB, we give in the column {\em Unit} the GRBM units used to derive the 
parameters reported in the catalog. The first number of the column identifies, 
for each GRB, the GRBM unit used for the determination of $T_{\rm 90}$,
$T_{\rm det}$ and
$T_{\rm a}$. This unit provided the light curve with the best signal-to-noise ratio. 
The second number gives the GRBM unit used for the determination of 
the GRB fluence, peak flux and spectral hardness $\Gamma$. When it 
was possible to improve the statistical quality of the results, the signal from
another unit was added in (third number). These two units have the lowest angular 
distance from the GRB.

\subsection{GRB Duration}

We report two different estimates of the GRB time duration in the 40--700
keV energy band. 
The first one is the usual $T_{\rm 90}$ as defined by the BATSE team
\citep{Kouveliotou93}. The technique adopted is described by
\citet{Koshut96} and it applies to the background subtracted
light curves. This method is applied to both  1~s 
and 7.8125 ms light curves. For the $T_{\rm 90}$ estimate, we have used 
the most illuminated unit. 

The second estimate of the GRB duration is $T_{\rm det}$, that gives
the time elapsed from the earliest GRB onset above a $2\sigma$ threshold
to its final disappearance below $2\sigma$. In the case of long GRBs ($\ge 2$~s),
the error associated to $T_{\rm det}$ 
was given by $R/\sqrt{2}$, where $R$ is the rebinning time listed in the column
{\em R} of Table~\ref{t:cat},
while in the case of short GRBs, this error was assumed to be equal to $R$. 
In addition to $T_{\rm det}$, for each GRB
we report in the catalog the integrated time $T_{\rm a}$ in which the burst 
was visible above a $2\sigma$ level and the number $N_{\rm a}$ of intervals in which
the $2\sigma$ level was exceeded in the time profile. The error associated 
to $T_{\rm a}$ was determined
assuming an uncertainty of $R/\sqrt{2}$ for each of the $N_{\rm a}$ intervals. 
From $T_{\rm a}$ an estimate of the integrated quiescent time $T_q$ of each GRB can be derived. 

For short GRBs with available high time resolution data (112 events), 
the $T_{\rm det}$ estimate was obtained from the 7.8125~ms light curves. For
the $T_{\rm a}$ determination, the $2\sigma$ level was derived using the Poissonian
distribution, i.e. determining the count threshold $n_{th}$ such that, 
if the count rate $n$ is higher than $n_{th}$, the probability that this 
count excess is due to a Poissonian fluctuation is lower than 5\%.

\subsection{Fluence and spectral hardness}

Using the GRBM response function discussed above, and the XSPEC software package,
for each GRB, 
its 40--700 keV fluence has been derived from the 
2 channel background--subtracted spectra (40-700 keV and $>$ 100 keV) 
integrated over $T_{\rm det}$ with the assumption of a power--law ({\sc pl}) spectrum. 
We refer to these spectra as ``2--channel spectra''. 
The power--law spectral index $\Gamma$ derived is reported in the column {\em $\Gamma$}
of the catalog. 
The {\sc pl} model was a forced choice for those GRBs 
for which the 240--channel spectra (integrated over 128 s) did not provide 
useful results. Given that the {\sc pl} choice could introduce a
systematic error in the fluence estimate, we corrected the fluences derived from
the 2--channel spectra in the following way. Using the strong GRBs (163) 
for which both the 2--channel  and 240--channel spectra could be used, 
we compared the fluence estimates derived from the 2--channel spectra 
with those derived from the dead--time corrected 240--channel spectra, in which 
the best input spectral model was adopted.
Thus we derived an average correction factor $CF$, that was used to correct the
2--channel fluences. This factor was found to approximately show a Gaussian distribution 
with a $\sigma \sim 0.01$ and no significant dependence on GRB fluence.

\subsection{Peak Flux}

For the 40--700 keV peak flux estimate, we adopted the same procedure followed for
the fluence estimate. Using the GRBM response function and assuming a {\sc pl} model,
for long GRBs, we deconvolved the 40--700 keV the peak counts over the minimum time
interval around the peak that gave a significant peak flux estimate. For long
GRBs, in general, we used the 1~s background--subtracted light curves in the 
two (40-700~keV and $>$100~keV) energy channels. 
However, for those GRBs for which the 1~s light curves provided a peak count
rate with a signal-to-noise ratio lower than 3, a rebinning of the data
was performed. The rebinning factor R is given in the last column of the
catalog (a value of 1 corresponds to 1 s light curves).
 
For short GRBs,  we first derived the 40--700 keV peak flux $F_p^{1s}$ from the 1~s 
light curves as above. Given that a 1~s integration time is too long, we corrected 
$F_p^{1s}$ for the ratio $R_C$ between the 125 ms 40--700 keV peak count rate
$C_p^{125ms}$ normalized to a 1 ~s time and the corresponding 1~s peak count rate 
$C_p^{1s}$ adopted for the deconvolution of the two--channel spectra. The reported
40--700 keV peak flux is given by $F_p = F_p^{1s} \times R_C$.

\section{Results of statistical analysis of the catalog}

We performed a statistical analysis of the data reported in the 
GRBM Catalog,  for checking their mutual consistency, for
comparing our results with those obtained with BATSE, and for getting
new results.

\subsection{Consistency tests}
\label{s:tests}

\begin{figure}
\begin{center}
\includegraphics[angle=270,width=0.90\textwidth]{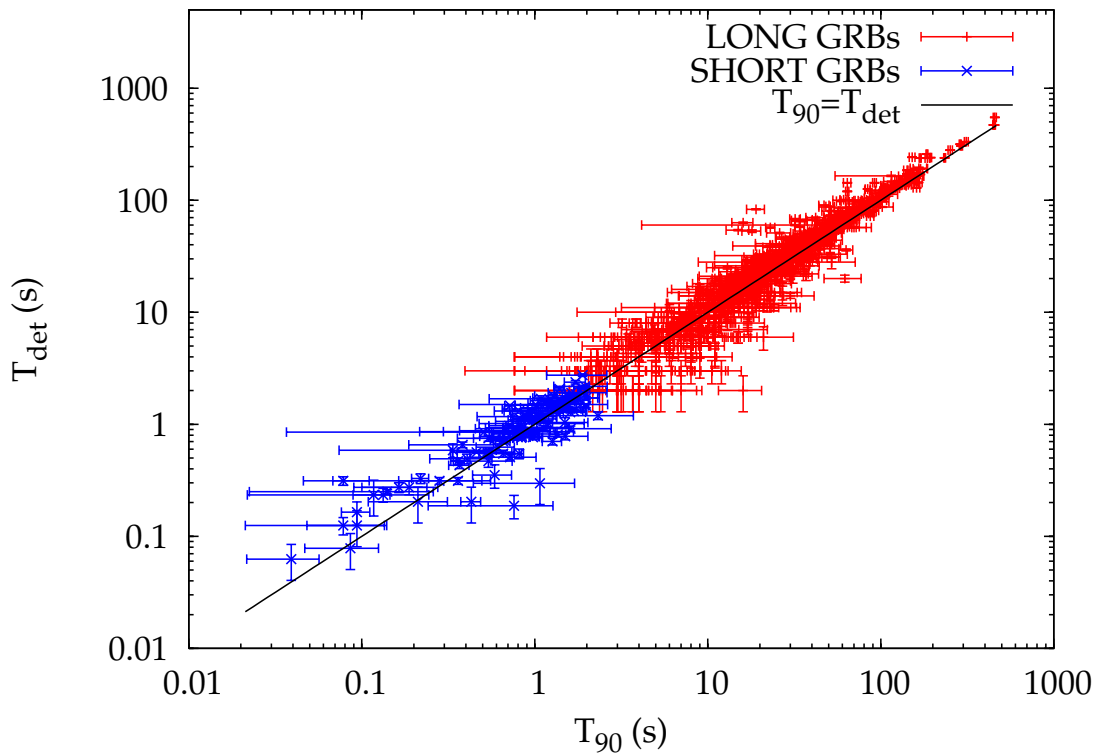}
\end{center}
\caption{Correlation between $T_{\rm 90}$ and $T_{\rm det}$ GRB time durations. Short
GRBs in blue color. The central continuous line gives the  curve $T_{\rm
90} = T_{\rm det}$ .
} 
\label{f:tdet_vs_t90}
\end{figure}

We tested the dependence of the  $T_{\rm det}$ estimate on $T_{\rm 90}$. The result
is shown in Fig.~\ref{f:tdet_vs_t90}. We find that, within statistical uncertainties,
$T_{\rm det}$ is fully correlated with $T_{\rm 90}$, as expected. In principle, we would expect
to get $T_{\rm det}$ equal or longer than $T_{\rm 90}$. However, for some weak GRBs, this 
expectation is not satisfied, mainly due to the larger uncertainty in the determination
of the $T_{\rm det}$ estimate with respect to that of $T_{\rm 90}$.

\begin{figure}
\begin{center}
\includegraphics[angle=270,width=0.4\textwidth]{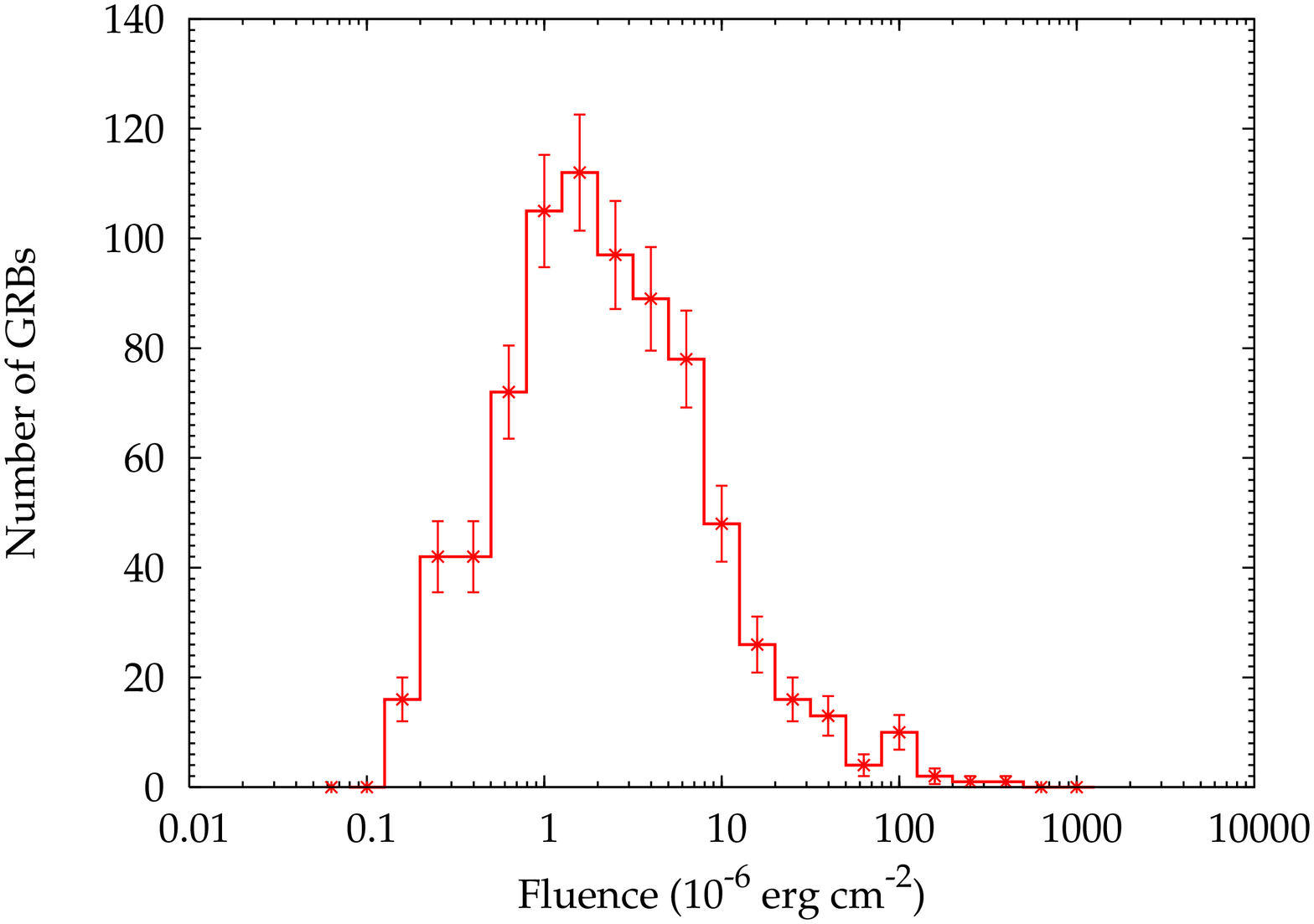}%
\includegraphics[angle=270,width=0.4\textwidth]{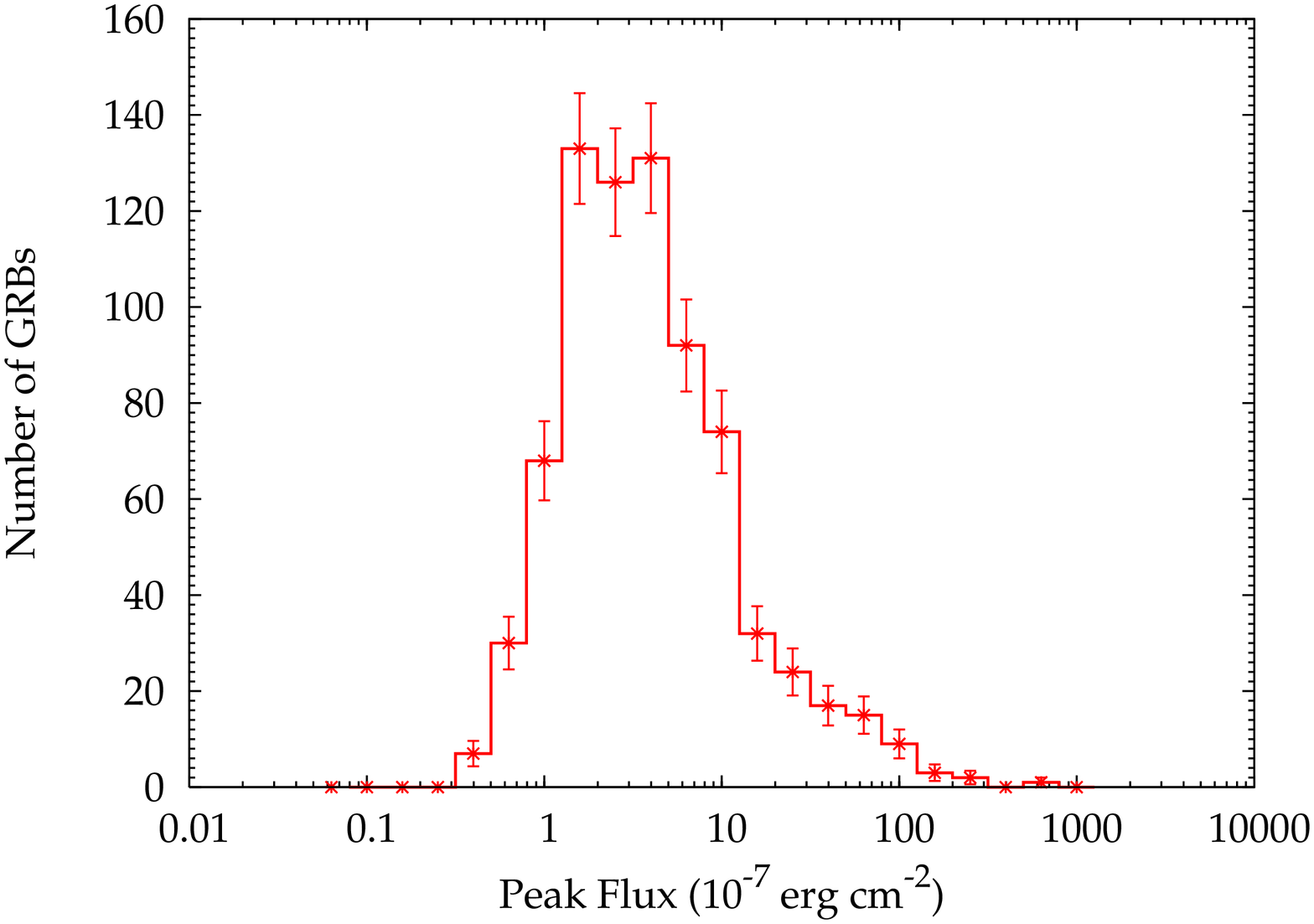}
\end{center}
\caption{Distribution of fluence ({\em left}) and peak flux ({\em right})  of the GRBs detected 
with the \sax\ GRBM.
} 
\label{f:peak-flux_fluence}
\end{figure}

\begin{figure}
\begin{center}
\includegraphics[angle=270,width=0.90\textwidth]{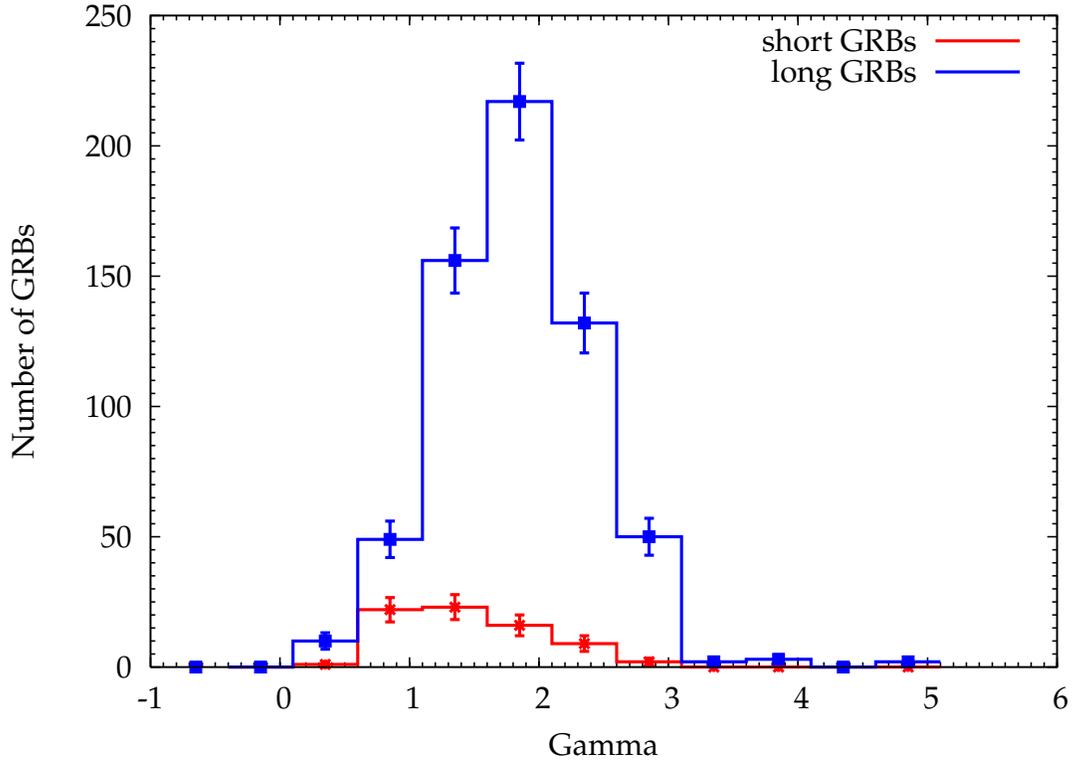}
\end{center}
\caption{Distribution of the GRB spectral hardness for short and long GRBs. The GRB hardness is
 defined as the index $\Gamma$ of the of the best power--law fit to the 2--channel spectra.   
} 
\label{f:distr_Gamma}
\end{figure}

The distributions of the GRB peak flux and fluence are shown in Fig.~\ref{f:peak-flux_fluence}.
As can be seen, both are clearly skewed, as expected as  a consequence of the 
GRBM sensitivity threshold. We derive from these Figures the limit
sensitivity of the GRBM: it is $\sim 4\times 10^{-8}$~erg~cm$^{-2}$~s$^{-1}$  
in terms of peak flux , while it is about $\sim 2\times 10^{-7}$~erg~cm$^{-2}$ in terms of fluence.
Also the hardness distribution for short and long GRBs has been derived and shown 
in Fig.~\ref{f:distr_Gamma}. As can be seen the distribution derived for long GRBs is almost symmetrical 
with a mean value $\Gamma_{long} \simeq2$ and a saturation around  
$\Gamma \simeq 3$, likely due to the limited sensitivity of the instrument 
at high photon energies. {Instead that derived for short GRBs is asymmetrical with a positive
skewness and a mean value of $\Gamma_{short} \simeq1.5$. At a significance level of $3.2\times 10^{-5}$, 
we find that of the $\Gamma$  mean value the long GRBs is higher than that of short GRBs}.

\subsection{Comparison with the BATSE results}

\begin{figure}
\begin{center}
\includegraphics[angle=270,width=0.90\textwidth]{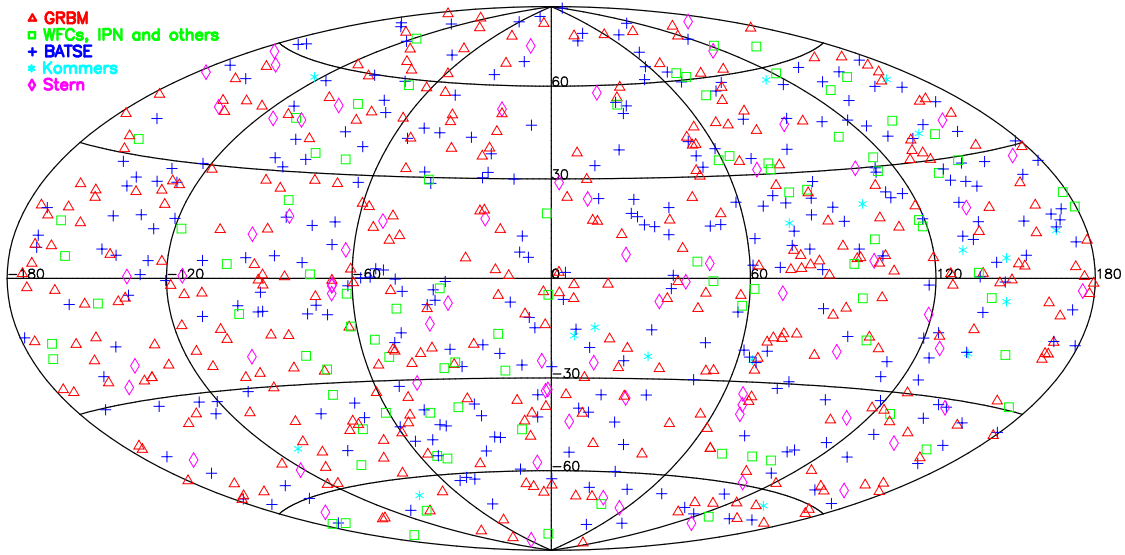}
\end{center}
\caption{Sky distribution (in Galactic coordinates) of GRBs detected with the \sax\ GRBM. 
In different colors
and symbols the GRB also found in other catalogs found with other instrumentation.
Also the GRBs identified with the \sax\ WFCs are shown.
} 
\label{f:sky_distr}
\end{figure}

The first test performed is the spatial distribution of the detected GRBs in the sky. 
The result is shown in Fig.~\ref{f:sky_distr}. We confirm the
isotropical distribution of the GRBs as found by BATSE \citep{Fishman94,Paciesas99}.
In Table~\ref{t:angular} we report the corresponding statistical results for 
detecting a Galactic association, i.e., our measures of the dipole and quadrupole 
moments \citep{Briggs93} in Galactic and equatorial coordinates.

\begin{figure}
\begin{center}
\includegraphics[angle=270,width=0.4\textwidth]{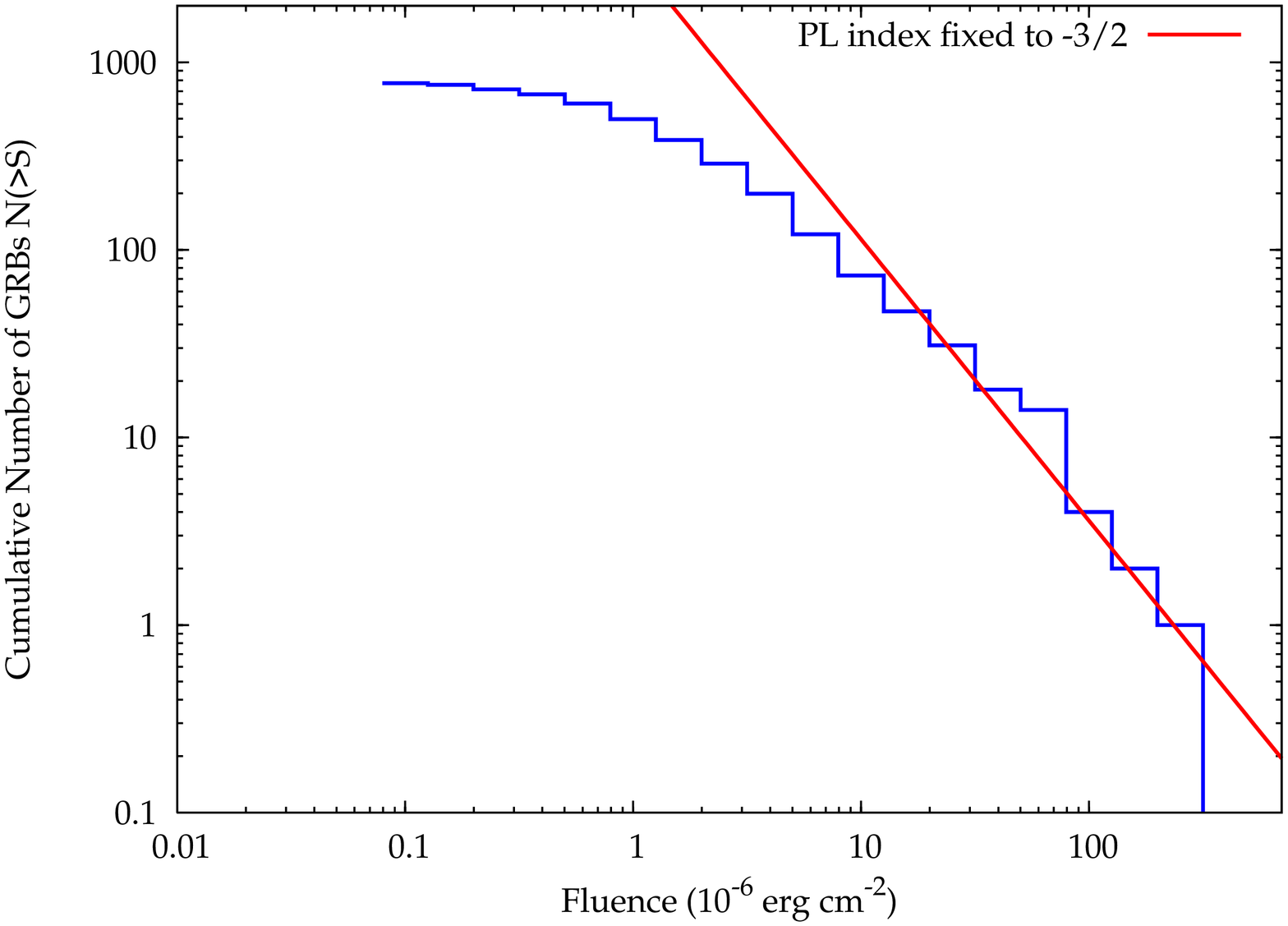}%
\includegraphics[angle=270,width=0.4\textwidth]{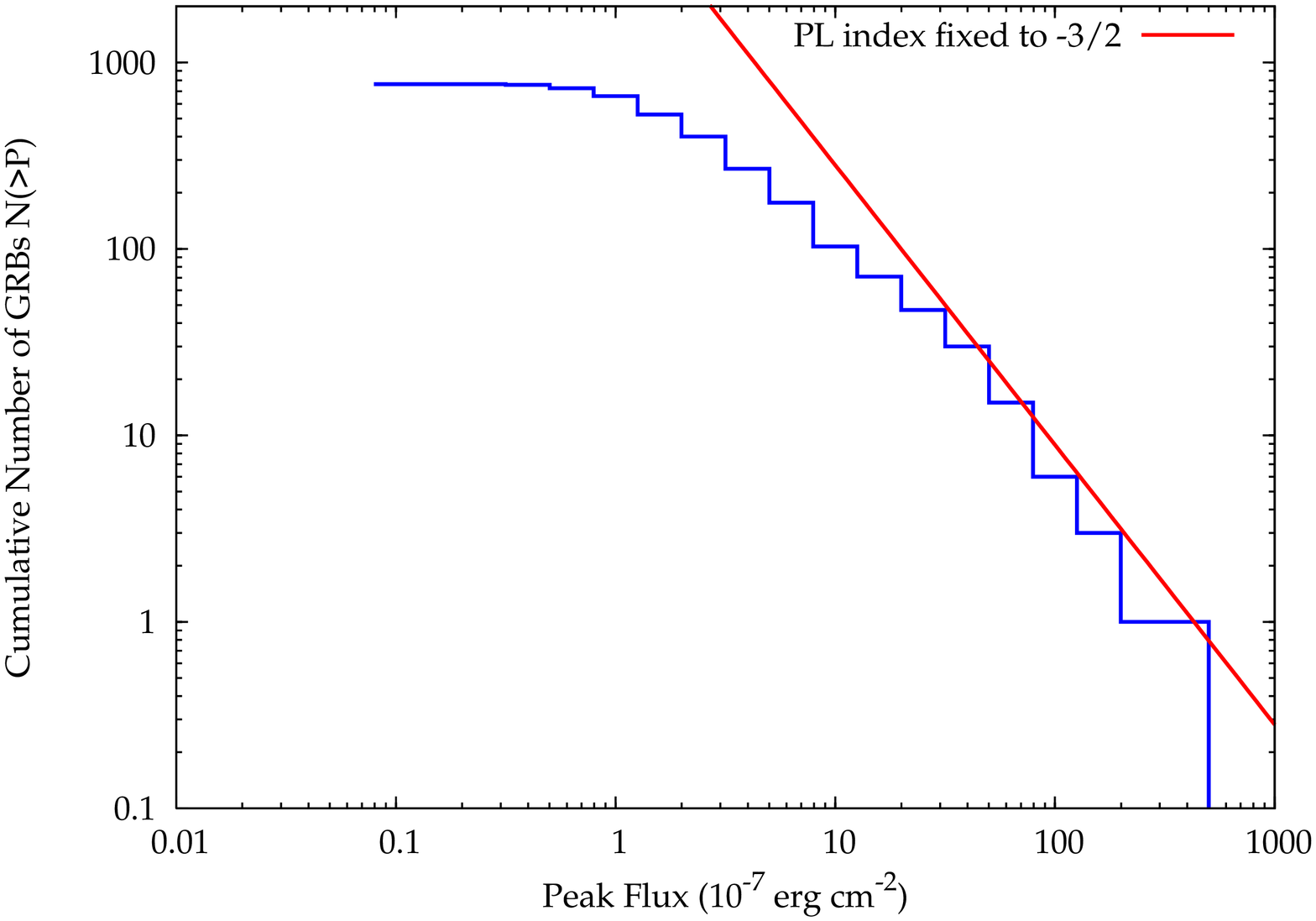}
\end{center}
\caption{Cumulative distribution of the 40--700 keV fluence (left panel) and peak flux (right panel)
of the GRBs detected with the \sax\ GRBM. The continuous line gives the 
distribution (a power--law of index $-3/2$) that is expected in the case the observed GRBs are 
homogeneously distributed in an Eucledian space throughout the sampled volume.   
} 
\label{f:logN}
\end{figure}

\begin{figure}
\begin{center}
\includegraphics[angle=0,width=0.4\textwidth]{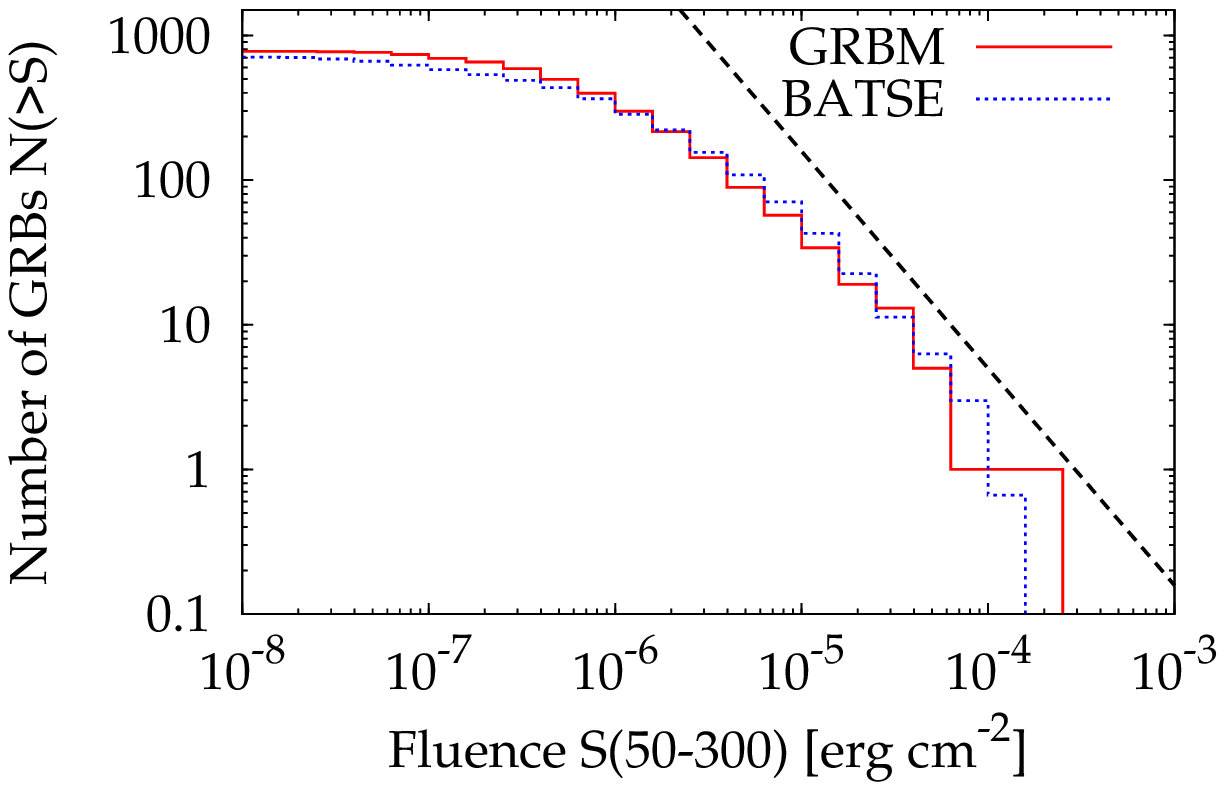}%
\includegraphics[angle=0,width=0.4\textwidth]{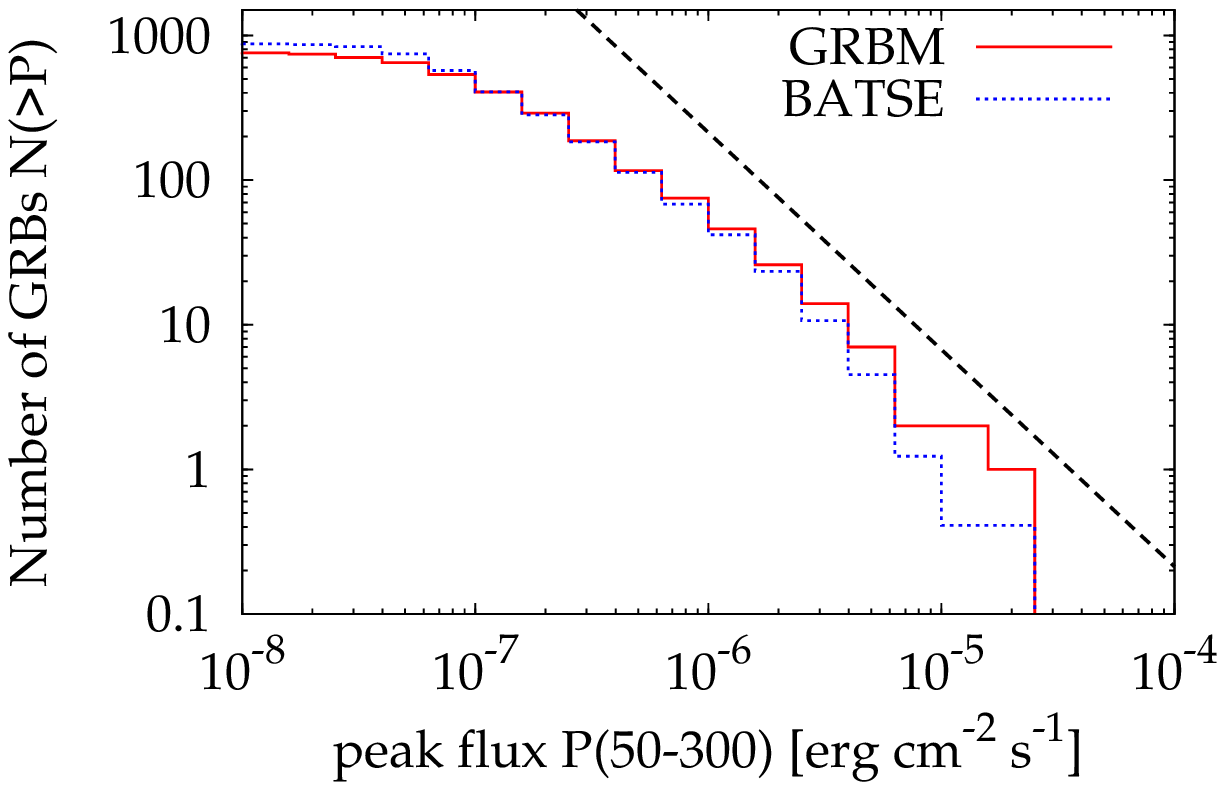}
\end{center}
\caption{Comparison of the \sax\ GRBM and BATSE 50--300 keV fluence (left panel) and peak flux (right
panel) cumulative distributions. The continuous line gives the 
distribution (a power--law of index $-3/2$) that is expected in the case the observed GRBs are 
homogeneously distributed in an Eucledian space throughout the sampled volume.   
} 
\label{f:logN-BATSE}
\end{figure}

We also derived the $\log N$--$\log S$ and $\log N$--$\log P$ distributions, 
where $N$ is the number of GRBs with  the 40--700 keV  fluence $S$ (or {40--700 keV} peak flux $P$) 
higher than given values. The results are shown in Fig.~\ref{f:logN}.  To compare our results 
with those derived with BATSE  in the 50--300 keV energy range  
(e.g. Paciesas et al. 1999\nocite{Paciesas99}), we first estimated the GRB fluence and peak flux 
in the 50--300 keV using the derived 
power--law spectral index $\Gamma$ . The derived distributions compared with those obtained
with BATSE are shown in Fig.~\ref{f:logN-BATSE}. Using a Kolmogorov-Smirnov (KS) test, we find that
at a significance level of 1\% the GRBM and BATSE $\log N$--$\log S$ distributions are consistent for 
fluences $> 1\times 10^{-6}$~\ergcm\,  while at a significance level of 5\% the  
$\log N$--$\log P$ distributions are consistent with each other for 
peak fluxes $>1\times 10^{-7}$~\ergcms.

\begin{figure}
\begin{center}
\includegraphics[angle=270,width=0.9\textwidth]{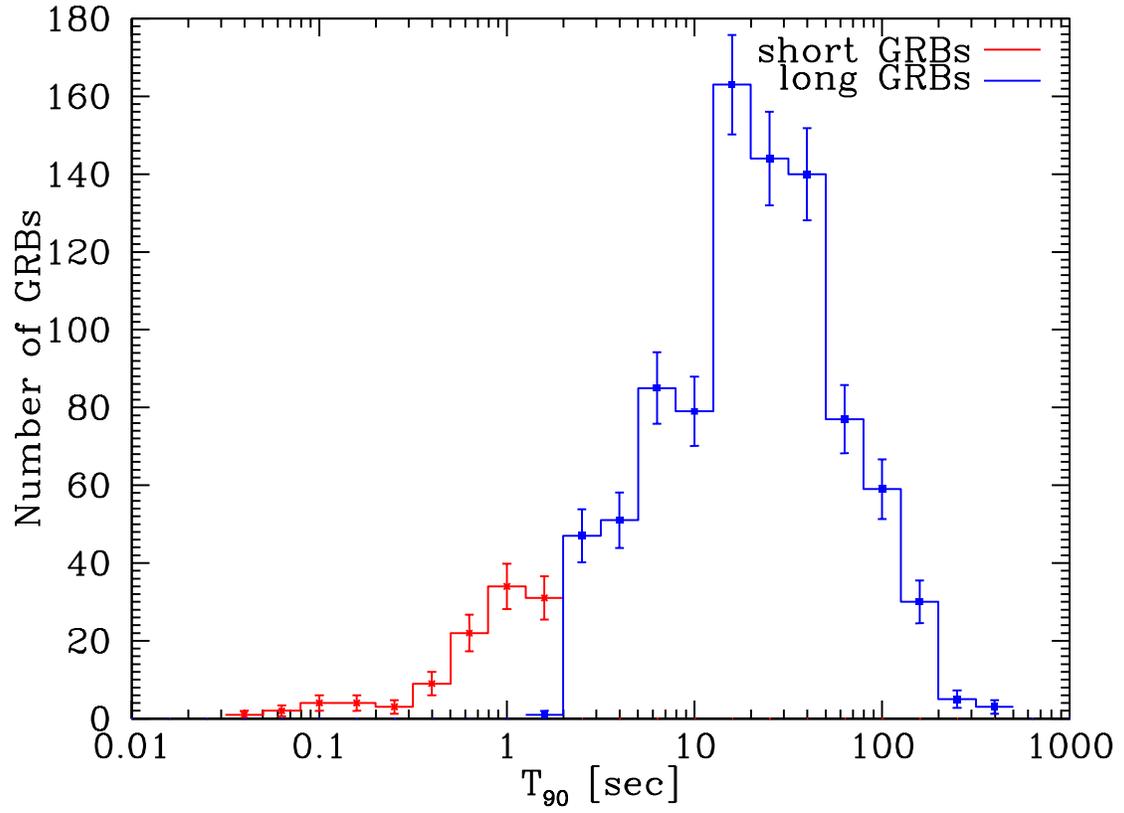}
\end{center}
\caption{Distribution of the GRBM GRBs according to their $T_{\rm 90}$ duration. Only
short GRB triggered by the on board logic were included. 
{\em In red color}: short GRBs; {\em in blue color}: long GRBs.   
} 
\label{f:distr_t90_tdet}
\end{figure}

The distributions of the GRBM GRBs with  $T_{\rm 90}$ and $T_{\rm det}$ have also been
investigated. Both distributions are similar. The results for $T_{\rm 90}$
are shown in Fig.~\ref{f:distr_t90_tdet}. As can be seen,
the bimodal behavior found with BATSE (see, e.g., Kouveliotou et al. 1993
\nocite{Kouveliotou93}) is confirmed, even if the distribution of
$T_{\rm 90}$ for short GRBs appears less pronounced and displaced toward higher durations with respect
to BATSE. This discrepancy is due to the lower efficiency of our trigger system
to short GRBs, which, for almost the entire mission duration, used 1~s as short integration
time (see Table~\ref{t:settings}). This bias does not allow to test the presence 
of a third class of GRBs as claimed by \citet{Horvath08}.  

\begin{figure}
\begin{center}
\includegraphics[angle=270,width=0.90\textwidth]{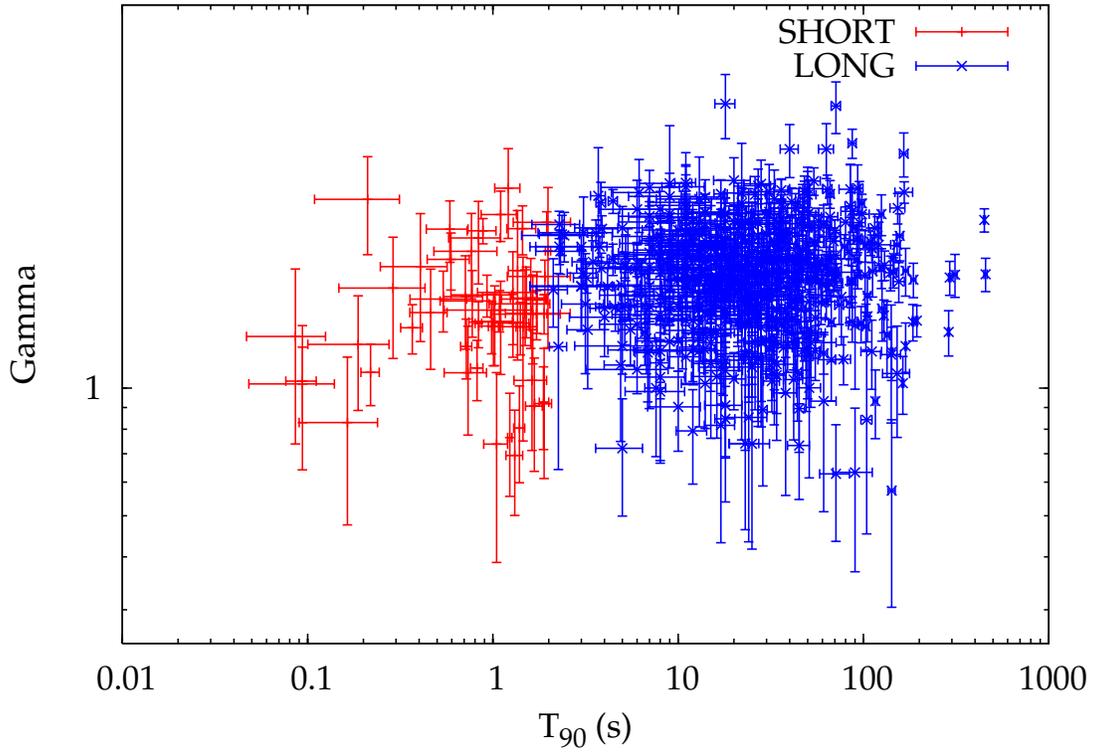}
\end{center}
\caption{Dependence of the GRB hardness on $T_{\rm 90}$ duration. As hardness parameter
 we use the power--law photon index $\Gamma$ listed in the Catalog. 
{\em In red color}: short GRBs; {\em in blue color}: long GRBs.   
} 
\label{f:hardness_vs_t90}
\end{figure}

We have also investigated the dependence of the GRB hardness $\Gamma$ as a function of 
the time duration $T_{\rm 90}$. The result is shown
in Fig.~\ref{f:hardness_vs_t90}. While it is apparent that there is no correlation of $\Gamma$ with the
GRB time duration within either the long or the short GRBs, using the nonparametric Spearman 
and Kendall correlation tests, we find a slight correlation (significance level of 2\%)   
between $\Gamma$ and GRB duration 
when both short and long GRBs are taken into account. This result confirms the result found
using the $\Gamma$ distribution of long and short GRBs (see section~\ref{s:tests}).

\subsection{Other tests}

In addition to the previous tests, other  tests have been derived from our catalog, 
thanks to the determination of $T_{\rm det}$ and of the activity time $T_{\rm a}$. 

\begin{figure}
\begin{center}
\includegraphics[angle=270,width=0.9\textwidth]{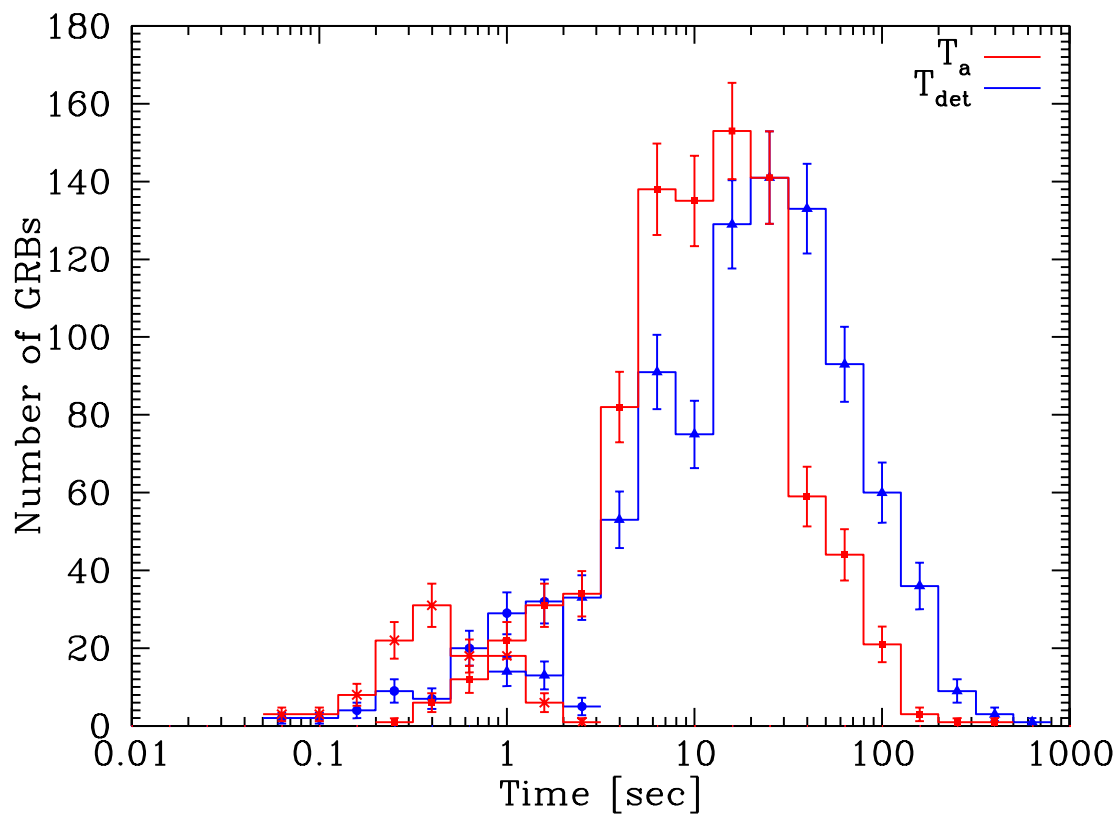}
\end{center}
\caption{Distribution of the integrated active emission time $T_{\rm a}$ ({\em red}) compared 
with the total duration $T_{\rm det}$ ({\em blue}). {\em Symbols for
$T_{\rm det}$}: filled triangles
for long GRBs, filled circles for short GRBs. {\em Symbols for $T_{\rm a}$}: filled squares for
long GRBs, stars for short GRBs.} 
\label{f:ta}
\end{figure}

In Fig.~\ref{f:ta} we show the distribution of $T_{\rm a}$ compared with that
of $T_{\rm det}$.
As can be seen, the maximum
activity time detected is about 200 s, whereas we find GRBs with $T_{\rm det}$ up to 600 s. 

\begin{figure}
\begin{center}
\includegraphics[angle=270,width=0.49\textwidth]{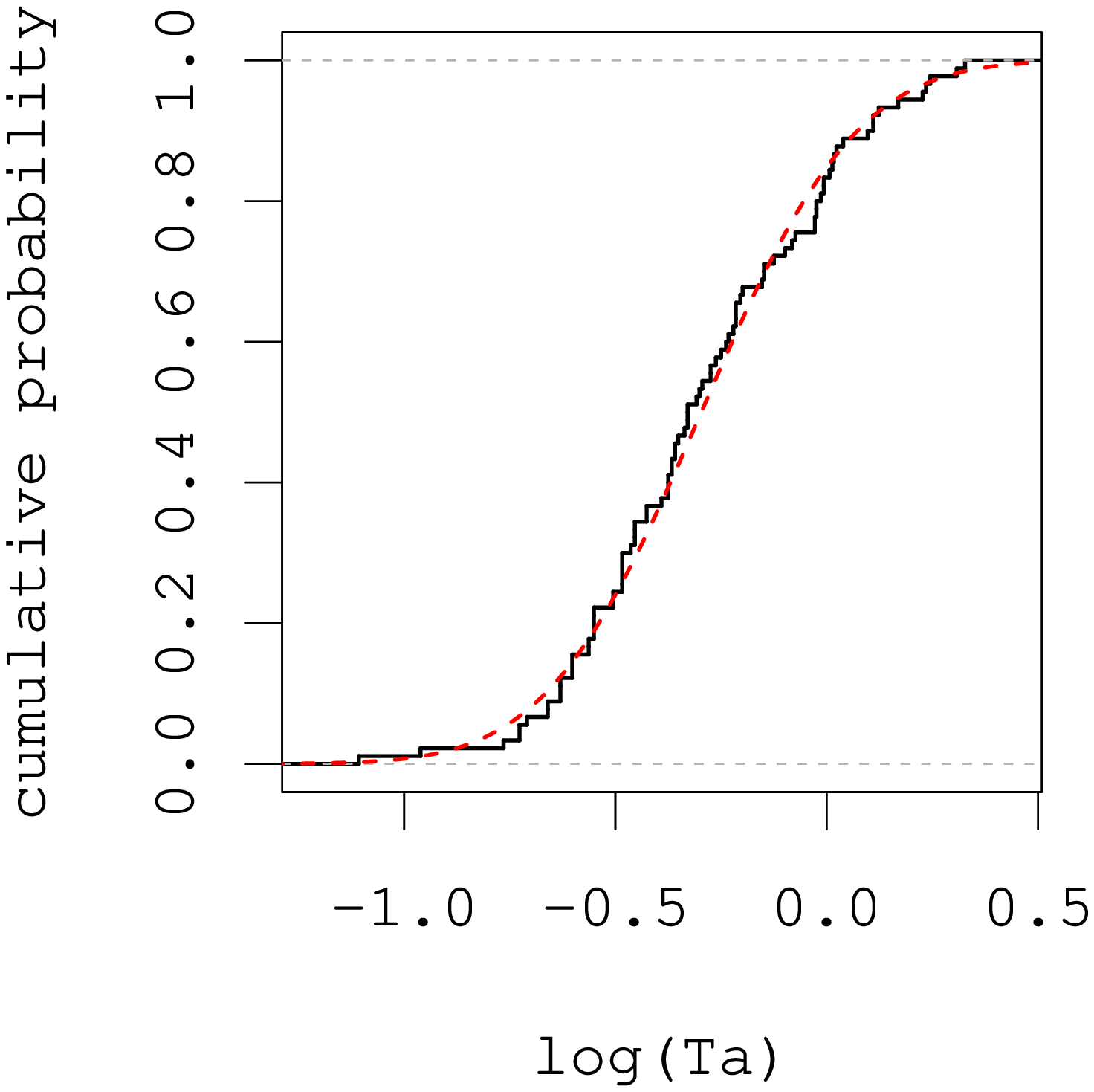}%
\includegraphics[angle=270,width=0.49\textwidth]{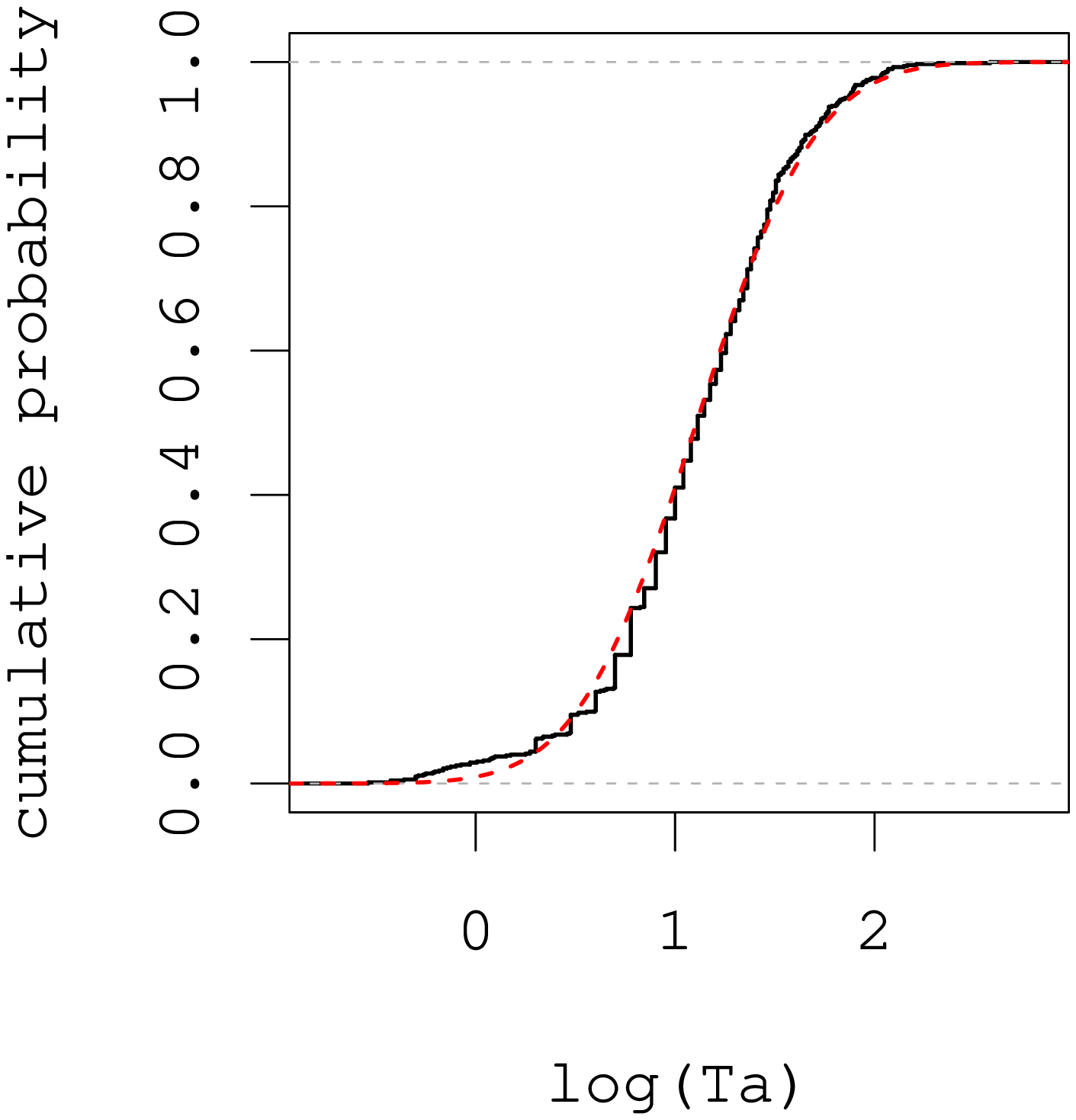}
\end{center}
\caption{Cumulative distribution of the active time $T_{\rm a}$ compared with that expected
from a log-normal distribution (dashed line). {\em Left panel}: short GRBs;
{\em right panel}: long GRBs.} 
\label{f:tact}
\end{figure}

\begin{figure}
\begin{center}
\includegraphics[angle=270,width=0.49\textwidth]{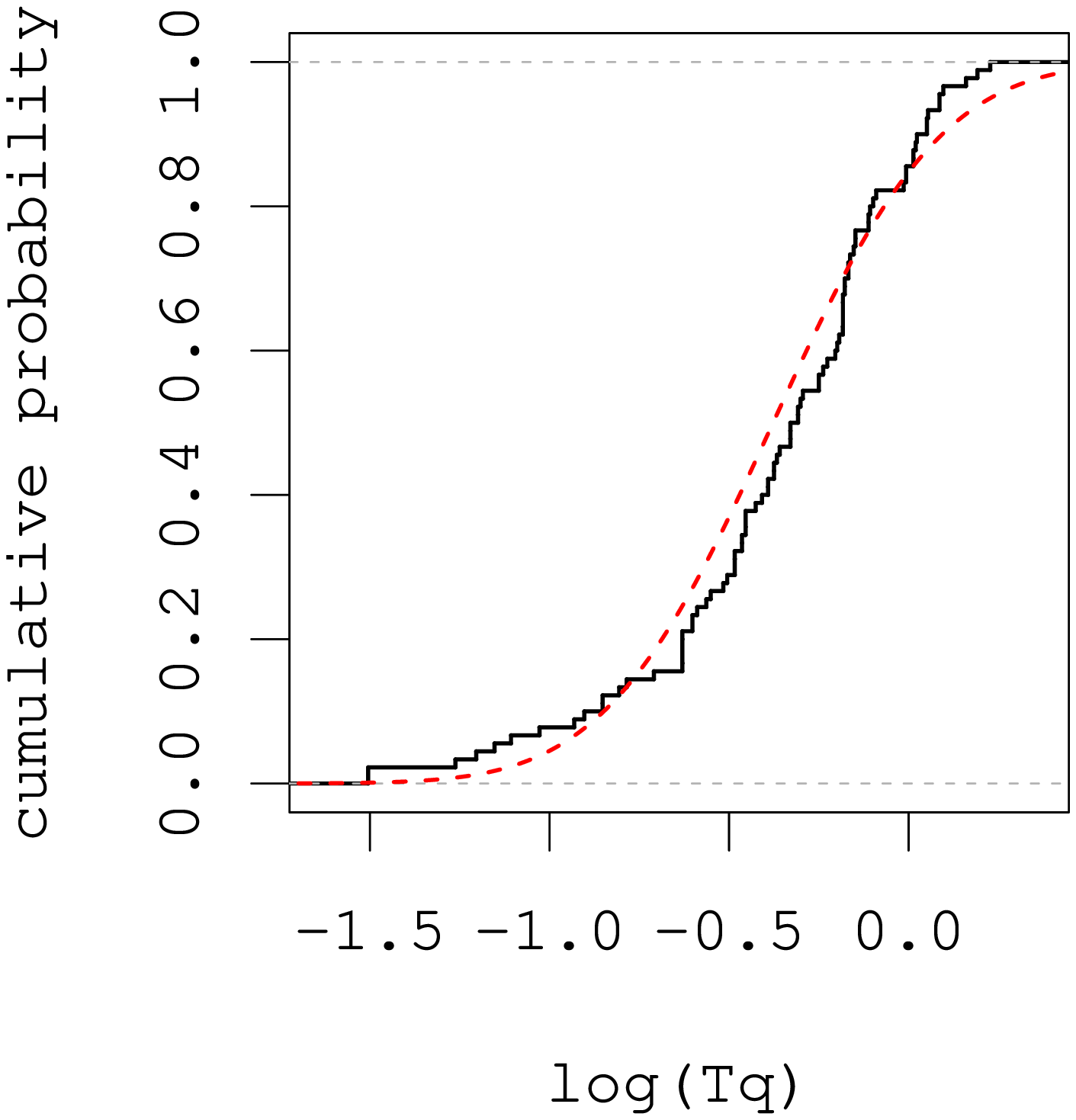}%
\includegraphics[angle=270,width=0.49\textwidth]{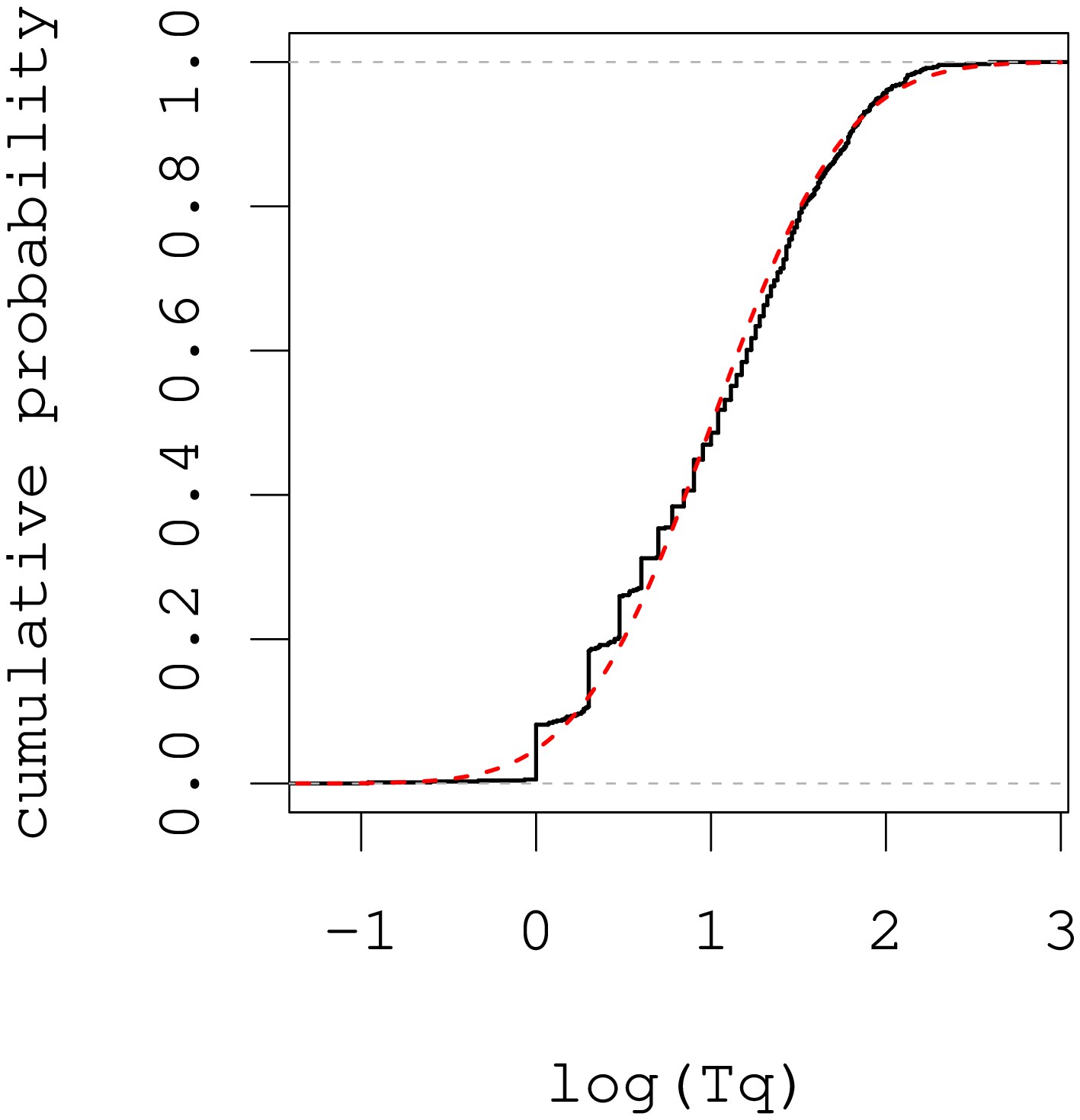}
\end{center}
\caption{Cumulative distribution of the quiescent time $T_q$ compared with that expected
from a log-normal distribution (dashed line). {\em Left panel}: short GRBs;
{\em right panel}: long GRBs.} 
\label{f:tq}
\end{figure}

In Fig.~\ref{f:tact} we show the cumulative distribution of the integrated active time $T_{\rm a}$. 
As can be seen from this figure and as found from a Shapiro--Wilk test \citep{Shapiro65}, 
for short GRBs, this distribution is consistent with a log-normal distribution ($p$--value 
of 0.58) in the case of short GRBs, while it is inconsistent with the latter ($p$--value 
of $4.6\times 10^{-8}$ ) in the case of long GRBs. 
In Fig.~\ref{f:tq} we show the cumulative distribution of the integrated quiescent time
$T_q$ obtained by subtracting $T_{\rm a}$ from $T_{\rm det}$. As can be seen from this figure
and from the Shapiro--Wilk test, for either short and long GRBs this distribution is 
inconsistent with a log-normal, with a lower probability chance 
for long GRBs ($p$--value of $1.9\times 10^{-4}$ in the case of short GRBs against
a $p$--value of $7.5\times 10^{-8}$ in the case of long GRBs).

\section{Discussion}

The first complete catalog of the GRBs detected with the \sax\ GRBM includes 1082 GRBs,
for most of which we have reported complete information (position, duration, peak flux,
fluence and spectral hardness).

Concerning the GRB positions, we have reported the most accurate coordinates available
for each GRBs included in the catalog. In the next future, for some of them, more accurate
(or the first determination of) positions will be available from the Inter--Planetary 
Network (K. Hurley et al., in preparation), allowing a more accurate determination (or the first
determination) of the fluence and peak flux values.

Some relevant results obtained with the BATSE catalogs (e.g., isotropical distribution 
of GRBs in the sky, deviation of the log$N$--log$S$ and log$N$--log$P$ cumulative 
distributions from that expected in the case of an homogeneous distribution of GRBs 
in an Euclidean space, two--peak distribution of the GRBs with time duration, 
higher hardness of the short GRBs) are confirmed. 

For the first time, our catalog includes, in addition to the classical
$T_{\rm 90}$ GRB 
duration, also the total duration $T_{\rm det}$ and the integrated time $T_{\rm a}$ during
which the GRB intensity exceeds the 2$\sigma$ level. We find that the GRB active time
has a cutoff at about 200 s, in spite that we find GRBs with $T_{\rm det}$ up to 600 s.
This result can have a relation with the physics of the unknown engine that gives rise
to the GRBs. It seems that this engine has a maximum active time of 200 s. 
The integrated active is found consistent with a log-normal distribution for either
short and long GRBs.

The origin of quiescent times is still unknown. The quiescent time properties have 
been investigated by various authors \citep{Nakar02,Drago07}. We limit our investigation to 
the active time $T_{\rm a}$ and quiescent time $T_q$, both integrated over the entire GRB duration. 
In the case of $T_{\rm a}$, we find that its distribution is consistent with a log-normal distribution
for short GRBs, while it is inconsistent with the latter in the case of long GRBs. 
In the case of the integrated $T_q$, we find that the cumulative distribution is  inconsistent with
a log-normal distribution in the case of short and long GRBs.
The consequences of these results would require further discussion, which is beyond the scope of this paper.
We limit us to state that models of GRB engines and of their behavior with time should 
take into account these statistical results.

\begin{acknowledgments}

Many people contributed to the success of the GRBM 
instrument. We wish to thank all of them. The \sax\ mission was a joint program of the Italian 
space agency ASI and of the Netherlands Agency for Aerospace Programs. 
This research was supported by the Ministry of Education, University and Research
of Italy (PRIN 2005-025417 devoted to GRBs). 

\end{acknowledgments}

\bibliographystyle{apj}
\bibliography{apj-jour,grb_ref}

\hyphenation{Post-Script Sprin-ger}
\begin{thebibliography}{31}
\expandafter\ifx\csname natexlab\endcsname\relax\def\natexlab#1{#1}\fi

\bibitem[{{Amati} {et~al.}(1997){Amati}, {Cinti}, {Feroci}, {Costa},
  {Frontera}, {dal Fiume}, {Collina}, {Nicastro}, {Orlandini}, {Palazzi},
  {Rapisarda}, \& {Zavattini}}]{Amati97}
{Amati}, L., {Cinti}, M.~N., {Feroci}, M., {Costa}, E., {Frontera}, F., {dal
  Fiume}, D., {Collina}, P., {Nicastro}, L., {Orlandini}, M., {Palazzi}, E.,
  {Rapisarda}, M., \& {Zavattini}, G. 1997, in Proc. SPIE Vol. 3114, p.
  176-185, EUV, X-Ray, and Gamma-Ray Instrumentation for Astronomy VIII, Oswald
  H. Siegmund; Mark A. Gummin; Eds., ed. O.~H. {Siegmund} \& M.~A. {Gummin},
  176--185

\bibitem[{{Boella} {et~al.}(1997){Boella}, {Chiappetti}, {Conti}, {Cusumano},
  {del Sordo}, {La Rosa}, {Maccarone}, {Mineo}, {Molendi}, {Re}, {Sacco}, \&
  {Tripiciano}}]{Boella97a}
{Boella}, G., {Chiappetti}, L., {Conti}, G., {Cusumano}, G., {del Sordo}, S.,
  {La Rosa}, G., {Maccarone}, M.~C., {Mineo}, T., {Molendi}, S., {Re}, S.,
  {Sacco}, B., \& {Tripiciano}, M. 1997, \aaps, 122, 327

\bibitem[{{Briggs}(1993)}]{Briggs93}
{Briggs}, M.~S. 1993, \apj, 407, 126

\bibitem[{{Brown}(2002)}]{Brown02}
{Brown}, F. e.~a. 2002, {Trans. Am. Nucl. Soc.}, 87, 273

\bibitem[{{Calura} {et~al.}(2000){Calura}, {Rapisarda}, {Frontera},
  {Montanari}, {Guidorzi}, {Amati}, {Feroci}, {Costa}, \& {Collina}}]{Calura00}
{Calura}, F., {Rapisarda}, M., {Frontera}, F., {Montanari}, E., {Guidorzi}, C.,
  {Amati}, L., {Feroci}, M., {Costa}, E., \& {Collina}, P. 2000, in AIP Conf.
  Proc. 526: Gamma-ray Bursts, 5th Huntsville Symposium, ed. R.~M. {Kippen},
  R.~S. {Mallozzi}, \& G.~J. {Fishman}, 721

\bibitem[{{Costa} {et~al.}(1998){Costa}, {Frontera}, {dal Fiume}, {Amati},
  {Cinti}, {Collina}, {Feroci}, {Nicastro}, {Orlandini}, {Palazzi},
  {Rapisarda}, \& {Zavattini}}]{Costa98}
{Costa}, E., {Frontera}, F., {dal Fiume}, D., {Amati}, L., {Cinti}, M.~N.,
  {Collina}, P., {Feroci}, M., {Nicastro}, L., {Orlandini}, M., {Palazzi}, E.,
  {Rapisarda}, M., \& {Zavattini}, G. 1998, Advances in Space Research, 22,
  1129

\bibitem[{{Costa} {et~al.}(1997){Costa}, {Frontera}, {Heise}, {Feroci}, {in 't
  Zand}, {Fiore}, {Cinti}, {dal Fiume}, {Nicastro}, {Orlandini}, {Palazzi},
  {Rapisarda}, {Zavattini}, {Jager}, {Parmar}, {Owens}, {Molendi}, {Cusumano},
  {Maccarone}, {Giarrusso}, {Coletta}, {Antonelli}, {Giommi}, {Muller}, {Piro},
  \& {Butler}}]{Costa97}
{Costa}, E., {Frontera}, F., {Heise}, J., {Feroci}, M., {in 't Zand}, J.,
  {Fiore}, F., {Cinti}, M.~N., {dal Fiume}, D., {Nicastro}, L., {Orlandini},
  M., {Palazzi}, E., {Rapisarda}, M., {Zavattini}, G., {Jager}, R., {Parmar},
  A., {Owens}, A., {Molendi}, S., {Cusumano}, G., {Maccarone}, M.~C.,
  {Giarrusso}, S., {Coletta}, A., {Antonelli}, L.~A., {Giommi}, P., {Muller},
  J.~M., {Piro}, L., \& {Butler}, R.~C. 1997, \nat, 387, 783

\bibitem[{{Drago} \& {Pagliara}(2007)}]{Drago07}
{Drago}, A., \& {Pagliara}, G. 2007, \apj, 665, 1227

\bibitem[{{Feroci} {et~al.}(1997){Feroci}, {Frontera}, {Costa}, {dal Fiume},
  {Amati}, {Bruca}, {Cinti}, {Coletta}, {Collina}, {Guidorzi}, {Nicastro},
  {Orlandini}, {Palazzi}, {Rapisarda}, {Zavattini}, \& {Butler}}]{Feroci97}
{Feroci}, M., {Frontera}, F., {Costa}, E., {dal Fiume}, D., {Amati}, L.,
  {Bruca}, L., {Cinti}, M.~N., {Coletta}, A., {Collina}, P., {Guidorzi}, C.,
  {Nicastro}, L., {Orlandini}, M., {Palazzi}, E., {Rapisarda}, M., {Zavattini},
  G., \& {Butler}, R.~C. 1997, in Proc. SPIE Vol. 3114, p. 186-197, EUV, X-Ray,
  and Gamma-Ray Instrumentation for Astronomy VIII, Oswald H. Siegmund; Mark A.
  Gummin; Eds., ed. O.~H. {Siegmund} \& M.~A. {Gummin}, 186--197

\bibitem[{{Fishman} {et~al.}(1994){Fishman}, {Meegan}, {Wilson}, {Brock},
  {Horack}, {Kouveliotou}, {Howard}, {Paciesas}, {Briggs}, {Pendleton},
  {Koshut}, {Mallozzi}, {Stollberg}, \& {Lestrade}}]{Fishman94}
{Fishman}, G.~J., {Meegan}, C.~A., {Wilson}, R.~B., {Brock}, M.~N., {Horack},
  J.~M., {Kouveliotou}, C., {Howard}, S., {Paciesas}, W.~S., {Briggs}, M.~S.,
  {Pendleton}, G.~N., {Koshut}, T.~M., {Mallozzi}, R.~S., {Stollberg}, M., \&
  {Lestrade}, J.~P. 1994, \apjs, 92, 229

\bibitem[{{Frail} {et~al.}(1997){Frail}, {Kulkarni}, {Nicastro}, {Feroci}, \&
  {Taylor}}]{Frail97}
{Frail}, D.~A., {Kulkarni}, S.~R., {Nicastro}, L., {Feroci}, M., \& {Taylor},
  G.~B. 1997, \nat, 389, 261

\bibitem[{{Frontera}(2004)}]{Frontera04c}
{Frontera}, F. 2004, in Astronomical Society of the Pacific Conference Series,
  ed. M.~{Feroci}, F.~{Frontera}, N.~{Masetti}, \& L.~{Piro}, 3--11

\bibitem[{{Frontera} {et~al.}(1997){Frontera}, {Costa}, {dal Fiume}, {Feroci},
  {Nicastro}, {Orlandini}, {Palazzi}, \& {Zavattini}}]{Frontera97}
{Frontera}, F., {Costa}, E., {dal Fiume}, D., {Feroci}, M., {Nicastro}, L.,
  {Orlandini}, M., {Palazzi}, E., \& {Zavattini}, G. 1997, \aaps, 122, 357

\bibitem[{{Guidorzi}(2002)}]{Guidorzi02}
{Guidorzi}, C. 2002, PhD thesis, Physics Department, University of Ferrara

\bibitem[{{Guidorzi} {et~al.}(1998){Guidorzi}, {Amati}, {Feroci}, {Costa},
  {Frontera}, {dal Fiume}, \& {Orlandini}}]{Guidorzi98}
{Guidorzi}, C., {Amati}, L., {Feroci}, M., {Costa}, E., {Frontera}, F., {dal
  Fiume}, D., \& {Orlandini}, M. 1998, in The Active X-ray Sky: Results from
  BeppoSAX and RXTE, ed. L.~{Scarsi}, H.~{Bradt}, P.~{Giommi}, \& F.~{Fiore},
  664

\bibitem[{{Guidorzi} {et~al.}(2001){Guidorzi}, {Frontera}, {Montanari}, F., ,
  {Amati}, {Costa}, \& M.}]{Guidorzi01}
{Guidorzi}, C., {Frontera}, F., {Montanari}, E., F., C., , {Amati}, L.,
  {Costa}, E., \& M., F. 2001, in Gamma Ray Bursts in the Afterglow Era, ed.
  E.~{Costa}, F.~{Frontera}, \& J.~{Hjorth}, 43

\bibitem[{{Horv{\'a}th} {et~al.}(2008){Horv{\'a}th}, {Bal{\'a}zs}, {Bagoly}, \&
  {Veres}}]{Horvath08}
{Horv{\'a}th}, I., {Bal{\'a}zs}, L.~G., {Bagoly}, Z., \& {Veres}, P. 2008,
  \aap, 489, L1

\bibitem[{{Jager} {et~al.}(1997){Jager}, {Mels}, {Brinkman}, {Galama},
  {Goulooze}, {Heise}, {Lowes}, {Muller}, {Naber}, {Rook}, {Schuurhof},
  {Schuurmans}, \& {Wiersma}}]{Jager97}
{Jager}, R., {Mels}, W.~A., {Brinkman}, A.~C., {Galama}, M.~Y., {Goulooze}, H.,
  {Heise}, J., {Lowes}, P., {Muller}, J.~M., {Naber}, A., {Rook}, A.,
  {Schuurhof}, R., {Schuurmans}, J.~J., \& {Wiersma}, G. 1997, \aaps, 125, 557

\bibitem[{{Kaneko} {et~al.}(2006){Kaneko}, {Preece}, {Briggs}, {Paciesas},
  {Meegan}, \& {Band}}]{Kaneko07}
{Kaneko}, Y., {Preece}, R.~D., {Briggs}, M.~S., {Paciesas}, W.~S., {Meegan},
  C.~A., \& {Band}, D.~L. 2006, \apjs, 166, 298

\bibitem[{{Kommers} {et~al.}(2001){Kommers}, {Lewin}, {Kouveliotou}, {van
  Paradijs}, {Pendleton}, {Meegan}, \& {Fishman}}]{Kommers01}
{Kommers}, J.~M., {Lewin}, W.~H.~G., {Kouveliotou}, C., {van Paradijs}, J.,
  {Pendleton}, G.~N., {Meegan}, C.~A., \& {Fishman}, G.~J. 2001, \apjs, 134,
  385

\bibitem[{{Koshut} {et~al.}(1996){Koshut}, {Paciesas}, {Kouveliotou}, {van
  Paradijs}, {Pendleton}, {Fishman}, \& {Meegan}}]{Koshut96}
{Koshut}, T.~M., {Paciesas}, W.~S., {Kouveliotou}, C., {van Paradijs}, J.,
  {Pendleton}, G.~N., {Fishman}, G.~J., \& {Meegan}, C.~A. 1996, \apj, 463, 570

\bibitem[{{Kouveliotou} {et~al.}(1993){Kouveliotou}, {Meegan}, {Fishman},
  {Bhat}, {Briggs}, {Koshut}, {Paciesas}, \& {Pendleton}}]{Kouveliotou93}
{Kouveliotou}, C., {Meegan}, C.~A., {Fishman}, G.~J., {Bhat}, N.~P., {Briggs},
  M.~S., {Koshut}, T.~M., {Paciesas}, W.~S., \& {Pendleton}, G.~N. 1993, \apjl,
  413, L101

\bibitem[{{Meegan} {et~al.}(1996){Meegan}, {Pendleton}, {Briggs},
  {Kouveliotou}, {Koshut}, {Lestrade}, {Paciesas}, {McCollough}, {Brainerd},
  {Horack}, {Hakkila}, {Henze}, {Preece}, {Mallozzi}, \& {Fishman}}]{Meegan96}
{Meegan}, C.~A., {Pendleton}, G.~N., {Briggs}, M.~S., {Kouveliotou}, C.,
  {Koshut}, T.~M., {Lestrade}, J.~P., {Paciesas}, W.~S., {McCollough}, M.~L.,
  {Brainerd}, J.~J., {Horack}, J.~M., {Hakkila}, J., {Henze}, W., {Preece},
  R.~D., {Mallozzi}, R.~S., \& {Fishman}, G.~J. 1996, \apjs, 106, 65

\bibitem[{{Metzger} {et~al.}(1997){Metzger}, {Djorgovski}, {Kulkarni},
  {Steidel}, {Adelberger}, {Frail}, {Costa}, \& {Frontera}}]{Metzger97}
{Metzger}, M.~R., {Djorgovski}, S.~G., {Kulkarni}, S.~R., {Steidel}, C.~C.,
  {Adelberger}, K.~L., {Frail}, D.~A., {Costa}, E., \& {Frontera}, F. 1997,
  \nat, 387, 878

\bibitem[{{Nakar} \& {Piran}(2002)}]{Nakar02}
{Nakar}, E., \& {Piran}, T. 2002, \mnras, 331, 40

\bibitem[{{Paciesas} {et~al.}(1999){Paciesas}, {Meegan}, {Pendleton}, {Briggs},
  {Kouveliotou}, {Koshut}, {Lestrade}, {McCollough}, {Brainerd}, {Hakkila},
  {Henze}, {Preece}, {Connaughton}, {Kippen}, {Mallozzi}, {Fishman},
  {Richardson}, \& {Sahi}}]{Paciesas99}
{Paciesas}, W.~S., {Meegan}, C.~A., {Pendleton}, G.~N., {Briggs}, M.~S.,
  {Kouveliotou}, C., {Koshut}, T.~M., {Lestrade}, J.~P., {McCollough}, M.~L.,
  {Brainerd}, J.~J., {Hakkila}, J., {Henze}, W., {Preece}, R.~D.,
  {Connaughton}, V., {Kippen}, R.~M., {Mallozzi}, R.~S., {Fishman}, G.~J.,
  {Richardson}, G.~A., \& {Sahi}, M. 1999, \apjs, 122, 465

\bibitem[{{Rapisarda} {et~al.}(1997){Rapisarda}, {Amati}, {Cinti}, {Feroci},
  {Costa}, {Collina}, {Zavattini}, {Frontera}, {Nicastro}, {Orlandini},
  {Palazzi}, \& {dal Fiume}}]{Rapisarda97}
{Rapisarda}, M., {Amati}, L., {Cinti}, M.~N., {Feroci}, M., {Costa}, E.,
  {Collina}, P., {Zavattini}, G., {Frontera}, F., {Nicastro}, L., {Orlandini},
  M., {Palazzi}, E., \& {dal Fiume}, D. 1997, in Proc. SPIE Vol. 3114, p.
  198-205, EUV, X-Ray, and Gamma-Ray Instrumentation for Astronomy VIII, Oswald
  H. Siegmund; Mark A. Gummin; Eds., ed. O.~H. {Siegmund} \& M.~A. {Gummin},
  198--205

\bibitem[{{Shapiro} \& {Wilk}(1965)}]{Shapiro65}
{Shapiro}, S.~S., \& {Wilk}, M.~B. 1965, {Biometrika}, 52, 591

\bibitem[{{Stern} {et~al.}(2001){Stern}, {Tikhomirova}, {Kompaneets},
  {Svensson}, \& {Poutanen}}]{Stern01}
{Stern}, B.~E., {Tikhomirova}, Y., {Kompaneets}, D., {Svensson}, R., \&
  {Poutanen}, J. 2001, \apj, 563, 80

\bibitem[{{Toor} \& {Seward}(1974)}]{Toor74}
{Toor}, A., \& {Seward}, F.~D. 1974, \aj, 79, 995

\bibitem[{{van Paradijs} {et~al.}(1997){van Paradijs}, {Groot}, {Galama},
  {Kouveliotou}, {Strom}, {Telting}, {Rutten}, {Fishman}, {Meegan}, {Pettini},
  {Tanvir}, {Bloom}, {Pedersen}, {Nordgaard-Nielsen}, {Linden-Vornle},
  {Melnick}, {van der Steene}, {Bremer}, {Naber}, {Heise}, {in 't Zand},
  {Costa}, {Feroci}, {Piro}, {Frontera}, {Zavattini}, {Nicastro}, {Palazzi},
  {Bennet}, {Hanlon}, \& {Parmar}}]{Vanparadijs97}
{van Paradijs}, J., {Groot}, P.~J., {Galama}, T., {Kouveliotou}, C., {Strom},
  R.~G., {Telting}, J., {Rutten}, R.~G.~M., {Fishman}, G.~J., {Meegan}, C.~A.,
  {Pettini}, M., {Tanvir}, N., {Bloom}, J., {Pedersen}, H.,
  {Nordgaard-Nielsen}, H.~U., {Linden-Vornle}, M., {Melnick}, J., {van der
  Steene}, G., {Bremer}, M., {Naber}, R., {Heise}, J., {in 't Zand}, J.,
  {Costa}, E., {Feroci}, M., {Piro}, L., {Frontera}, F., {Zavattini}, G.,
  {Nicastro}, L., {Palazzi}, E., {Bennet}, K., {Hanlon}, L., \& {Parmar}, A.
  1997, \nat, 386, 686

\end{thebibliography}

\clearpage

%
%

\begin{deluxetable}{l|ccc|ccc}
\tablecaption{GRBM trigger parameter and energy threshold settings.}
\tablewidth{0pt}
\tablehead{Epoch & \multicolumn{3}{c|}{Trigger parameters} & \multicolumn{3}{c}{Energy thresholds} \\
    & LIT (s) & SIT (s) & $n_\sigma$ & LLT (keV) & ULT (keV) & ACT (keV)} 
\startdata
Mission start time & 128 & 4 		& 8   & 32.8 & 700 & 100   \\
30 October 1996      & 128 & 1 		& 4   & 42.5 & 700 & 100  \\
28 November 1996      & 32  & 1 		& 4	& 42.5 & 700 & 100  \\
January 2002       & 32  & 0.125  	& 4   & 42.5 & 700 & 100  \\
\enddata
\label{t:settings}
\end{deluxetable}

\clearpage

\clearpage
\pagestyle{empty}
\changepage{1in}{1in}{-0.75in}{-0.75in}{}{}{}{}{}
%
%
\begin{deluxetable}{lccrrrrrrcccccccccc}
\tabletypesize{\tiny}
\rotate
\tablecaption{\sax\ GRBM Catalogue}
\tablewidth{0pt}
\tablehead{
\colhead{GRB} & \colhead{Time} & \colhead{HR} & 
\colhead{R.A.} & \colhead{Dec} & \colhead{Err} &
\colhead{$l$} & \colhead{$b$} & 
\colhead{Elev} & \colhead{CAT} & 
\colhead{$T_{\rm det}$} & \colhead{$T_{\rm a}$} & 
\colhead{$N_{\rm a}$} & \colhead{$T_{\rm 90}$} &
\colhead{Fluence} & \colhead{F$_{\rm peak}$} & 
\colhead{$\Gamma$} & \colhead{Unit} & \colhead{R} \\
 & \colhead{(HH.MM.SS)} & & 
\colhead{($^\circ$)} & \colhead{($^\circ$)} &
\colhead{($^\circ$)} & \colhead{($^\circ$)} &
\colhead{($^\circ$)} & \colhead{($^\circ$)} & & 
\colhead{(s)} & \colhead{(s)} & & \colhead{(s)} & 
\colhead{($10^{-6}$ cgs)} &
\colhead{($10^{-7}$ cgs)} & & & \\
}
\startdata
960703A & 07.39.48 & N & 142.00 & -9.00 & 36.0 & 241.8678 & 28.9165 & \nodata & G & $11.0\pm 1.4$ & $9.0\pm 2.8$ & 3 & $12.0\pm 4.2$ & $2.34 \pm 0.28$ & $2.81 \pm 0.50$ & $2.30 \pm 0.29$ & 2 2 - & 2  \\
960703B & 13.42.54 & N & 4.62 & -7.75 & 1.9 & 99.4676 & -69.1067 & 72.0 & 4B\_5526 & $78.0\pm 0.7$ & $58.0\pm 2.1$ & 10 & $71.0\pm 3.2$ & $11.11 \pm 0.75$ & $9.54 \pm 0.74$ & $1.88 \pm 0.14$ & 1 1 4 & 1  \\
960703C & 18.10.41 & N & 330.40 & -54.79 & 1.9 & 339.0609 & -48.6887 & 61.1 & 4B\_5527 & $1.0\pm 0.7$ & $1.0\pm 0.7$ & 1 & $3.0\pm 1.4$ & $0.91 \pm 0.13$ & $6.04 \pm 0.89$ & $1.73 \pm 0.37$ & 3 1 4 & 1  \\
960707A & 10.16.41 & Y & 320.95 & 82.51 & 1.9 & 116.5625 & 22.3662 & 23.8 & 4B\_5530 & $6.0\pm 0.7$ & $6.0\pm 0.7$ & 1 & $4.45 \pm 0.25$ & $3.34 \pm 0.25$ & $9.09 \pm 0.76$ & $2.76 \pm 0.18$ & 1 1 - & 1  \\
960707B & 16.26.26 & N & 162.39 & -14.34 & 9.5 & 263.6772 & 39.0682 & 71.7 & 4B\_5532 & $97.0\pm 0.7$ & $35.0\pm 3.5$ & 22 & $90.0\pm 21.5$ & $5.25 \pm 0.52$ & $3.25 \pm 0.48$ & $0.63 \pm 0.26$ & 3 3 - & 1  \\
960720 & 11.36.53 & Y & 262.65 & 49.08 & 0.1 & 75.7380 & 33.0885 & 66.0 & W & $6.0\pm 0.7$ & $6.0\pm 0.7$ & 1 & $6.0\pm 2.2$ & $1.95 \pm 0.16$ & $8.45 \pm 0.77$ & $1.66 \pm 0.18$ & 1 1 - & 1  \\
960723A & 04.46.03 & N & 94.99 & 43.25 & 2.2 & 170.6861 & 12.9251 & 67.0 & 4B\_5551 & $64.0\pm 0.7$ & $39.0\pm 2.1$ & 11 & $54.0\pm 4.5$ & $6.63 \pm 0.58$ & $3.10 \pm 0.46$ & $3.08 \pm 0.23$ & 2 2 - & 1  \\
960723B & 07.07.46 & N & \nodata & \nodata & \nodata & \nodata & \nodata & \nodata & \nodata & $3.0\pm 0.7$ & $3.0\pm 0.7$ & 1 & $3.0\pm 2.2$ & \nodata & \nodata & \nodata & 2 - - & 1  \\
\enddata

\tablecomments{
{\em Column 1}: GRB name (yymmdd); 
{\em column 2}: Trigger time in UT (hh.mm.ss); 
{\em column 3}: high time resolution data available for the GRB: yes (y), not (n); 
{\em column 4}: Right ascension (epoch 2000.0); 
{\em column 5}: Declination (epoch 2000.0); 
{\em column 6}: error radius; 
{\em column 7}: Galactic longitude; 
{\em column 8}: Galactic latitude;
{\em column 9}: Elevation angle with respect to the Earth limb; 
{\em column 10}: instrument/catalog that provided the best GRB 
coordinates: \sax\ GRBM (G); \sax\ Wide Field Cameras (W); ASM aboard RXTE (A); 
BATSE 4B catalog (4B) \citep{Paciesas99}; HETE 2 (H); Kommers catalog (K) \citep{Kommers01}; 
Inter-Planetary Network (I); RXTE PCA (P); Stern catalog (S) \citep{Stern01}; 
{\em column 11}: time elapsed from the earliest GRB onset above $2\sigma$ threshold
 to its final disappearance below $2\sigma$ threshold;
{\em column 12}: integrated fraction of $T_{\rm det}$ in which the emission level is above $2\sigma$;
{\em column 13}: number of time intervals in which the emission level is above $2\sigma$;
{\em column 14}: Time duration defined by the BATSE team \citep{Kouveliotou93};
{\em column 15}: 40--700 keV fluence in units $10^{-6}$~erg~cm$^{-2}$; 
{\em column 16}: 40--700 keV peak flux in units of $10^{-7}$~erg~cm$^{-2}$~s$^{-1}$;  
{\em column 17}: best fit photon index of the 2 channel data; 
{\em column 18}: detection units used for the derived parameter values (see text); 
{\em column 19}: rebinning factor of the default light curve adopted to derive the reported parameters.
{\em Default light curves}: 1~s light curves in the case of long GRB ($T_{90}\ge 2$~s),
7.8125~ms in the case of short GRBs.
}
\label{t:cat}
\end{deluxetable}

\clearpage
\pagestyle{plaintop}
\changepage{-1in}{-1in}{0.75in}{0.75in}{}{}{}{}{} 
%
%
\begin{deluxetable}{l|c}
\tablecaption{GRBs statistics}
\tablewidth{0pt}
\tablehead{\colhead{GRB type} & \colhead{Number}}
\startdata
GRBs with all parameters determined & 764 \\
GRBs with only information  about duration & 266 \\
GRBs with gaps in the data & 42 \\
GRBs without $F_{\rm peak}$ because of low statistics & 10 \\
\hline
Total GRBs & 1082 
\enddata
\label{t:info}
\end{deluxetable}

%
%
\begin{deluxetable}{lllll}
\tablecaption{Dipole and quadrupole statistics for the GRBs  detected by the \sax\ GRBM with determined direction}
\tablewidth{0pt}
\tablehead{
\colhead{Tested} & \colhead{Moment} & \colhead{Coordinate} & \colhead{Measured}  & \colhead{Expected} \\
		\colhead{Moment} & \colhead{Type}  & \colhead{System}    & \colhead{Value} &  \colhead{value$^a$}
}
\startdata
$\langle \cos \theta \rangle$ & Dipole  & Galactic & $-0.015 \pm 0.020$ & $0.000 \pm 0.009$ \\
$\langle \sin \delta \rangle$ & Dipole  & Equatorial & $0.026 \pm 0.020$ & $0.000 \pm 0.009$ \\
$\langle \sin^2 b - 1/3 \rangle$ & Quadrupole  & Galactic & $-0.003 \pm 0.010$ & $-0.007 \pm 0.005$ \\
$\langle \sin^2 \delta - 1/3 \rangle$ & Quadrupole  & Equatorial & $0.038 \pm 0.010$ & $0.039 \pm 0.005$  
\enddata
\tablenotetext{a}{for an isotropic distribution and corrected for the sky exposure map}
\tablecomments{$\theta$ is the angle between the direction to the burst and 
the Galactic center, $b$ is the Galactic latitude, $\delta$ is GRB declination.
}
\label{t:angular}
\end{deluxetable}

\clearpage

%
%

%
%

%
%

%
%
%
%

%
%

%
%

%
%

%
%

%
%

%

%
%

%
%
%
%

\end{document}